# Ultrafast two-colour X-ray emission spectroscopy reveals excited state landscape in a base metal dyad

M. Nowakowski[1,†], M. Huber-Gedert[1,†], H. Elgabarty[1], J. Kubicki[2], A. Kertem[2], N. Lindner[2], D. Khakhulin,[3] F. Lima,[3] T.-K. Choi[3,4], M. Biednov[3], N. Piergies[5], P. Zalden[3], K. Kubicek[3], A. Rodriguez-Fernandez[3], M. Alaraby Salem[1],T. Kühne[1], W. Gawelda[2,6,7], M. Bauer[1*]

[1] Chemistry Department and Center for Sustainable Systems Design (CSSD), Faculty of Science, Paderborn University, Warburger Straße 100, 33098 Paderborn, Germany

[2] Faculty of Physics, Adam Mickiewicz University, Uniwersytetu Poznańskiego 2, Poznań, 61-614, Poland

[3] European X-Ray Free-Electron Laser Facility GmbH, Holzkoppel 4, 22869 Schenefeld, Germany

[4] PAL-XFEL, Jigok-ro 127-80, 37673 Pohang, Republic of Korea

[5] Institute of Nuclear Physics Polish Academy of Sciences, Kraków, 31-342, Poland.

[6] Departamento de Química, Universidad Autónoma de Madrid, Campus Cantoblanco, 28047 Madrid, Spain

[7] IMDEA Nanociencia, Calle Faraday 9, 28049 Madrid, Spain

**ABSTRACT:** Effective photoinduced charge transfer makes molecular bimetallic assemblies attractive for applications as active light induced proton reduction systems. For a more sustainable future, development of competitive base metal dyads is mandatory. However, the electron transfer mechanisms from the photosensitizer to the proton reduction catalyst in base metal dyads remain so far unexplored. We study a Fe-Co dyad that exhibits photocatalytic $H_2$ production activity using femtosecond X-ray emission spectroscopy, complemented by ultrafast optical spectroscopy and theoretical time-dependent DFT calculations, to understand the electronic and structural dynamics after photoexcitation and during the subsequent charge transfer process from the $Fe^{II}$ photosensitizer to the cobaloxime catalyst. Using this novel approach, the simultaneous measurement of the transient $K\alpha$ X-ray emission at the iron and cobalt K-edges in a two-colour experiment is enabled making it possible to correlate the excited state dynamics to the electron transfer processes. The methodology, therefore, provides a clear and direct spectroscopic evidence of the Fe→Co electron transfer responsible for the proton reduction activity.

## INTRODUCTION

$Fe^{II}$ complexes can operate as light-harvesting components in bimetallic molecular assemblies or dyads, to convert solar to chemical energy by ultrafast charge transfer (CT) to a second, catalyst metal for photocatalytic water splitting.[1] In terms of sustainability, the second metal should be abundant. Cobaloxime fulfils this requirement.[2–4] Despite the reported short lifetimes of metal-to-ligand charge transfer (MLCT) states in iron(II) photosensitizers, photocatalytic proton reduction activity was reported for $Fe^{II}$-$Co^{III}$ dyads.[5] However, its activity remains mysterious, as no charge transfer from the Fe to the Co center could be observed experimentally. Thus, rational improvement of Fe-Co dyads requires a radically different approach to understand the working principle. A major challenge are the ultrafast photophysics at the $Fe^{II}$ center,[6] and the difficulty to monitor the CT from the photosensitizer to the catalyst with element specificity and in real-time.[1,7,8] Upon photoexcitation the excited state dynamics in dyads can involve MLCT and ligand-to-metal charge transfer states (LMCT), metal-centred (MC) and ligand-mediated metal-to-metal charge-transfer states (M'MCT).[9,10] The de-excitation cascade is an interplay between MC and CT states, modulated by intramolecular vibrational energy dissipation, strong spin-orbit coupling such as intersystem crossings (ISC), and internal conversions (IC).[8,11] The fundamental principles guiding the properties are typically identified using laser spectroscopy.[12,13] Noble metal complexes exhibit long-lived CT states, which can easily be followed with optical spectroscopy, due to the associated intense absorption bands in UV-Vis range.[14] On the other hand, in most iron photosensitizers, the smaller ligand field splitting leads to an unfavoured energetic order of E(MLCT) > E(MC).[2] In addition, MC states are hardly accessible in UV-Vis spectral range.[15]

Contrary, X-ray emission spectroscopy (XES) is very sensitive to MC states due to the localized character of core levels. Both $K\alpha$ ($2p\rightarrow1s$) and $K\beta$ ($3p\rightarrow1s$) emission lines provide characteristic signatures of the multiplicity of the involved transient MC states.[16] For monomeric iron carbene photosensitizers, femtosecond XES could uniquely reveal details of the excited states.[13,16,17] The excited states of $[Fe(bmip)_2]^{2+}$ [bmip = 2,6-bis(3-methyl-imidaz-

ole-1-ylidine)-pyridine] show a complex branching pattern. One path is dominated by a long-lived ³MLCT, while the second includes a rapid hot MLCT* to ³MC transition-connected to bond oscillations in form of wavepacket dynamics[11,16,18] which are a stabilizing factor for long-lived ³MLCT states.[19] Ultrafast X-ray absorption near edge structure spectroscopy (XANES) on photoactive Fe-Co Prussian blue analogues revealed that a spin transition at the Co centre preceeded CT between the Fe and Co center.[20] More recently, the photoinduced M'MCT transition in a bimetallic Fe-Ru assembly was shown to have a critical impact on the solvent organization around the excited molecule.[21]

We demonstrate the unique potential of a two-colour X-ray emission spectroscopy (2C-XES) in photocatalysis research. It allows for simultaneous, ultrafast detection of the Fe and Co Kα XES in a [Fe-BL-Co] assembly of a heteroleptic Fe$^{II}$ photosensitizer with two different biscarbene-pyridine ligands (C^N^C) connected to a cobaloxime catalyst via a bridging ligand (BL).[5] The dynamics of the excited state decay are monitored at the Fe and Co sites to follow the departure of the charge from the photosensitizer and its arrival at the catalyst in real-time. As such, our experimental approach eliminates uncertainties related to the charge transfer event timescale and solves the puzzle about a possible charge transfer in base metal dyads.

## RESULTS

The dyad is synthesized combining a heteroleptic tetra-NHC Fe$^{II}$ photosensitizer [Fe-BL] coordinated by a 2,6-bis[3-(2,6-diisopropylphenyl)imidazol-2-ylidene]pyridine and a 2,6-bis(3-methyl-imidazol-2-ylidene)-4,4'-bipyridine ligand (BL) with a Co$^{III}$ cobaloxime catalyst, as presented in Fig. 1a.[5] A 4,4′-bipyridine (bpy) linker connects both metals with a distance of 11 Å.[22,23] The ground state optical absorption spectra of the dyad in comparison to the constituting components [Fe-BL] (red) and cobaloxime (Co(dmgH)$_2$Cl(py)=[Co], green) are shown in Fig. 1c. The top panel shows two absorption bands for the photosensitizer at 398 nm and 481 nm (Fig. 1c, top). By coordination of the cobaloxime in [Fe-BL-Co] (black), the UV-Vis spectrum of the dyad changes in a distinct manner: The 398 nm band remains unchanged, but the 481 nm band is shifted to 497 nm, and a new band appears at 444 nm (Fig. 1c, top). Cobaloxime itself shows only a weak absorption around 400 nm.

**Quantum chemical assignment of the optical absorption bands** In order to understand the properties of the electronic excited states of [Fe-BL-Co], the involved transitions must be identified. As a first step, we carefully benchmarked TPSSh/TDDFT excited-state calculations of the photosensitizer against CASSCF/NEVPT2 (SI, sec. 1b-c). The latter combination of a multireference wavefunction with a perturbative treatment of electron correlation accounts for both static and dynamic electron correlation effects, and has been shown to yield highly accurate results, but computational cost steeply increase with the number of correlated orbitals.[24] Both techniques reveal that the bands at 398 nm and 481 nm in [Fe-BL] (Fig.1c, top, red) are a mixture of MLCT transitions from Fe$^{II}$ to both the terminal and the BL (SI, sec. 1a-b,e). Together with the TPSSh/TDDFT computations of UV-Vis spectrum, this assignment is used to understand the absorption properties of [Fe-BL-Co] (SI, sec. 1c and 1f). The lower panel of Fig. 1c shows the TDDFT spectrum of the dyad. The experimentally observed 398 nm absorption is described by transitions a and b (SI, sec. 1c). Like in the photosensitizer, they are composed of MLCT transitions from iron to the terminal and bridging ligand. Additionally, the electron density is transferred from the Fe$^{II}$ to the Co$^{III}$ center along the bridging ligand in the form of an M'MCT transition (SI, sec. 1c). The donor-acceptor contributions to the latter are shown in Fig. 1b. The absorption at 497 nm is dominated by an MLCT transition to the bridging ligand together with a weak M'MCT contribution (transition c). The shift of the 497 nm band in the dyad spectrum compared to the photosensitizer spectrum (481 nm) is well reproduced by TDDFT (480.6 nm vs 446.0 nm, SI, sec. 1b-c). It is due to an increased charge transfer to the terminal pyridine ring (transition c in Fig. 1) and revealed by charge transfer components obtained for the dyad (SI, sec. 1f). Unfortunately, TDDFT could not resolve the 444 nm band in the dyad spectrum. After re-evaluation of the former interpretation[5], it is assumed that it is also present in the photosensitizer spectrum but overlaps with the 481 nm band.

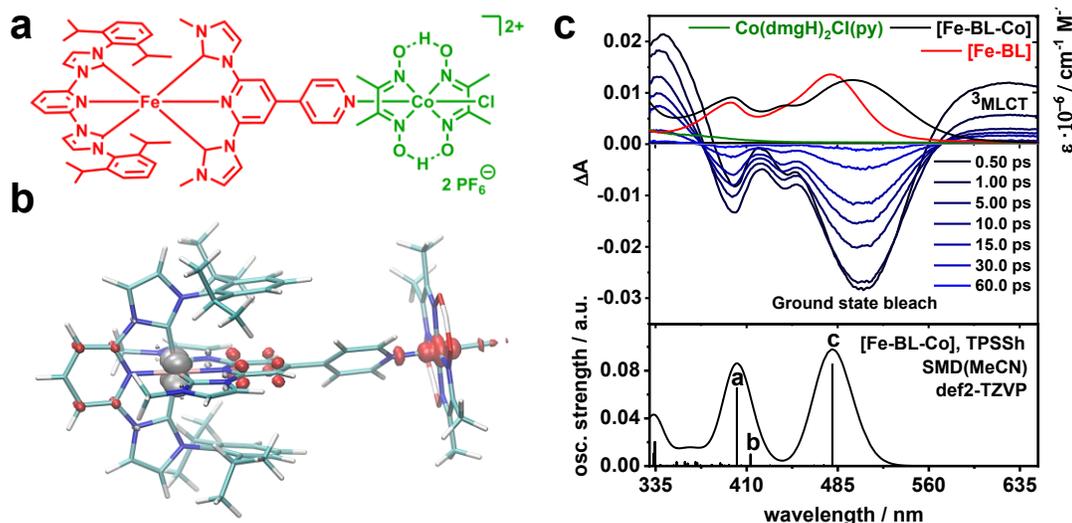

**Fig. 1. a)** Structure of the [Fe-BL-Co] dyad; **b)** Fe → Co CT (M'MCT): grey color indicates holes, red– electrons **c)** UV-Vis data for [Fe-BL], [Co] and [Fe-BL-Co] (top) overlapped with TAS results for selected delay times and TD-DFT UV-Vis spectrum for the dyad (bottom).

In conclusion, the nature of the transitions in the dyad is similar to those of the photosensitizer. Most important however, additional M'MCT contributions are found in all bands. Due to the unchanged absorption band at 398 nm in both compounds, only weak LMCT absorption of cobaloxime[2,25] and the M'MCT contribution in [Fe-BL-Co], an excitation wavelength of 400 nm was chosen for ultrafast transient spectroscopy.

**Transient Absorption Spectroscopy** Transient absorption spectroscopy (TAS) results for [Fe-BL-Co] are presented in Fig. 1c. The ground state bleach occurs at 370-560 nm, an excited-state absorption is observed <370 nm and >560 nm. The transient absorption >560 nm is assigned to a $^3$MLCT state[5] and its kinetics are composed of three time constants (SI, sec. 2). The first component (<100 fs) in this model takes into account all coherent artefacts[26] and possible $^1$MLCT contribution. The second component ($\tau_2$=350 fs) can be ascribed to either the relaxation from the hot $^3$MLCT* to thermally relaxed $^3$MLCT[27,28], or to a $^1$MLCT → $^3$MLCT transition.[11,15–17] This is supported by the excited-state TDDFT, where the first acceptor state for the 400 nm excitation is a $^1$MLCT. The longest component can be assigned to the lifetime of the relaxed $^3$MLCT state.[6,27,29,30] Due to the very similar results obtained in each fit (SI, sec. 2), a reliable average value of 12.8±1.2 ps for the lifetime of the $^3$MLCT state is obtained. For the constituting photosensitizer [Fe-BL] a $^3$MLCT lifetime of 11.1±0.4 ps is found (SI, sec. 2). The increased $^3$MLCT lifetime in [Fe-BL-Co] is interpreted as an indirect signature of CT processes, leaking into the relaxation channel over the $^3$MLCT state. Yet, it does not provide unequivocal proof for a Fe→Co charge transfer due to a lack of direct spectroscopic signatures for altered charge densities at both Fe and Co cobalt centers. This gap can be closed by XES.

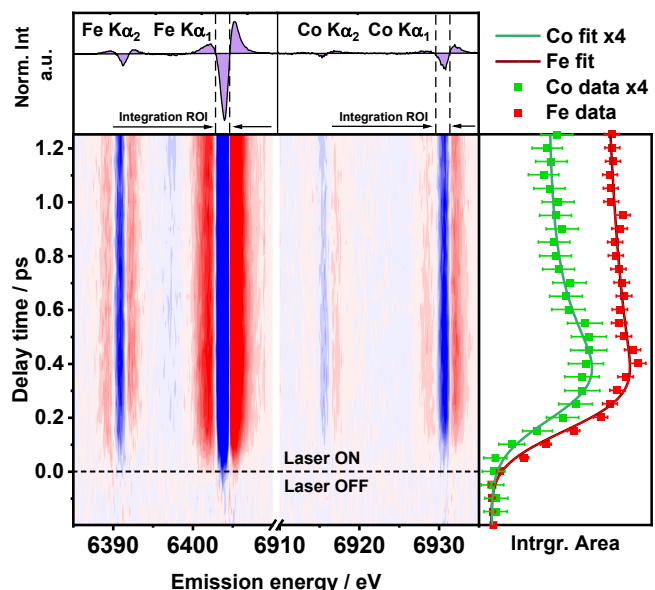

**Fig. 2.** Fe and Co K$\alpha_{1,2}$ transient XES line intensities of [Fe-BL-Co] for delay times of -0.2 – 1.25 ps. Top panel: transient XES signals at 1 ps delay time with integration regions of interest (ROIs) marked by vertical dashed lines. Right panel: integrated area under transient XES Fe K$\alpha_1$ and Co K$\alpha_1$ main feature in function of delay time (points) with corresponding fitted model (lines, top).

**Fe K$\alpha$ XES dynamics** XES is governed by different selection rules than the optical absorption. The XES signal originates from localized core electrons, and through the width of the K$\alpha_1$ XES line, it is directly proportional to the effective number of unpaired $d$-electrons[31] and covalency[32]. Using a von Hamos emission spectrometer,[33] in a 2C-XES scheme[34], spectra for both Fe and Co could be collected truly simultaneously, without any ambiguity of the time-zero on a femtosecond timescale.[35] Fig. 2 shows the early temporal evolution of the two XES signals and their kinetic traces along with the selected integration ranges for both elements.

In [Fe-BL-Co] three time constants of $\tau_{1,FeCo}$<0.14 ps, $\tau_{2,FeCo}$=10.38(40) ps and $\tau_{3,FeCo}$=1.74(18) ps are obtained from fitting of the transient kinetics at the iron K$\alpha_1$ emission, while for [Fe-BL] $\tau_{1,Fe}$~0.25 ps, $\tau_{2,Fe}$=8.98(27) ps and $\tau_{3,Fe}$=1.71(35) ps are found (SI, sec. 3). The most notable difference is thus the increased longest lifetime $\tau_2$ in the dyad, which is similarly observed in TA measurements.[5] The difference of 2 ps is attributed to the different sensitivity of TA and XES towards CT states - MC states are "optically silent" in UV-Vis spectral range.

Both singlet and triplet $^{1/3}$MLCT states of Fe compounds have near-identical K$\alpha$ XES signatures, since both have a single Fe-localized unpaired $d$-electron.[7] Moreover, since the coupling of the deep 2$p$ core-hole with the 3$d$ manifold is weak, K$\alpha$ XES has little sensitivity to ISC inside the MLCT manifold. With regards to metal spin multiplicity, there is a significant difference between the $^3$MLCT ($S_{loc}$=1/2) and $^3$MC ($S_{loc}$=1) states. Any relaxation process, involving either one of these states, to the singlet ground state is thus visible in transient Kα XES experiments. In agreement with previously reported values and the current TA results, the XES time constants are assigned in the following way: the shortest lifetime $\tau_1$ in both [Fe-BL] and [Fe-BL-Co] dyad corresponds to a $^3$MLCT*→$^3$MC channel.[36] The time constant $\tau_3$ can be attributed to the $^3$MC state decaying into the ground state.[37,38] The value of time constant $\tau_1$ fully agrees with reports of $^3$MLCT*→$^3$MC channels in Fe(II)-NHC complexes.[37,38] Moreover, TDDFT excited-state potential energy surfaces indeed identify a $^3$MC surface that intersects both the $^1$MLCT and the $^3$MLCT close to the Franck-Condon region. The $^3$MC is identified, using a Mulliken electron-hole population analysis, as a triplet state containing both a hole and an additional electron on the Fe metal, because of the Fe($d_{xy}/d_{yz}/d_{xz}$) → Fe($d_{x^2-y^2}/d_{z^2}$) transition (SI, sec. 1d). Time constant $\tau_2$ is assigned to the $^3$MLCT state.[36–38]

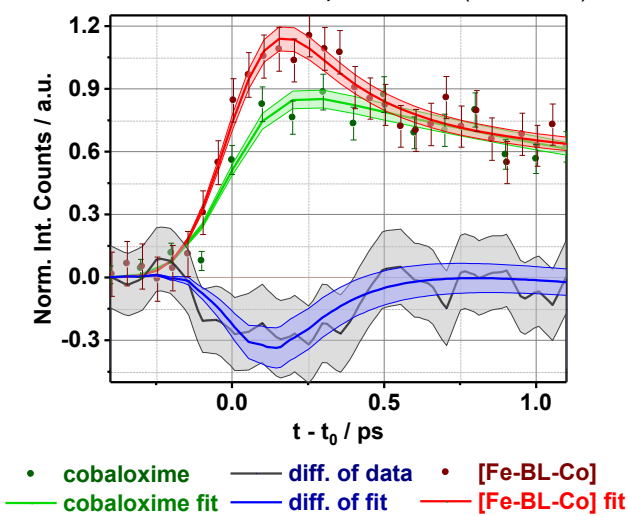

**Fig. 3.** Kinetics of the Co K$\alpha$ emission in [Fe-BL-Co] and cobaloxime after 400 nm excitation along with corresponding fits. Differential signal: [Co]- [Fe-BL-Co] is marked as blue lines (data+fit). Filled areas represent uncertainties.

**Co Kα XES dynamics** The Co Kα transient kinetics of the dyad [Fe-BL-Co] is constituted of three time constants $\tau_{1,FeCo}$=0.25(1) ps, $\tau_{2,FeCo}$=4.12(1.39) ps and $\tau_{3,FeCo}$~23.39 ps, a striking difference to pure cobaloxime [Co], where two time constants of $\tau_{1,Co}$=2.76(31) ps and $\tau_{2,Co}$=23.39(1.82) ps are found (SI, sec. 3a). This difference is also obvious in the kinetics of the Co Kα XES in [Co] (green) and [Fe-BL-Co] (red) in Fig. 3, which differ substantially in shape over the first 0.5 ps (SI, sec. 3a). The difference is caused by the short decay constant $\tau_{1,FeCo}$=0.25(1) ps that is not present in pure cobaloxime. This time constant thus represents a new excited state population channel, created by the formation of the dyad, which at later timescales is indispensable for photocatalytic hydrogen generation and for which the optical absorption data shows Fe-Co M'MCT contribution. Consequently, the differential signal in Fig. 3 is the real-time signature of CT from the Fe to the Co center in [Fe-BL-Co].

The direct excitation of the cobaloxime reflected in the Co kinetics of Fig. 3 (green) corresponds to an LMCT state. According to our DFT calculations,[2] the HOMO in cobaloxime is composed of degenerated π orbitals of the dmgH ligand, and the LUMO consists of the Co $d_{z^2}$ orbital leading to a very weak LMCT absorption at 396 nm. The observed low cross-section excitation populates this LMCT state of Co with a lifetime of 4.12(1.39) ps in [Fe-BL-Co].

Since cobaloxime has a documented activity as a proton reduction catalyst,[1,39] the increased catalytic activity of [Fe-BL-Co] compared to [Fe-BL] + cobaloxime originates from the M'MCT states in [Fe-BL-Co]. Note, the signal we observe originates from linear combination of differently excited species, since M'MCT and LMCT states cannot exist simultaneously in the same molecule. The result is evident, despite a low CT yield for our prototype dyad.

**Nuclear motion detected by Fe Kα XES** Both [Fe-BL] and [Fe-BL-Co] show a pronounced, but distinct, coherent nuclear wavepacket signatures in the transient iron Kα$_1$ kinetics. For the photosensitizer, the oscillations could be modelled by a single damped periodic function (Fig. 4, SI, sec. 3c), while in case of the dyad, it is composed of two contributions (Fig. 4 and SI, sec. 3c). [Fe-BL] and [Fe-BL-Co] share a half-period of 0.28(2) ps and 0.26(3) ps, respectively. Similar oscillation half-periods were observed in other systems.[16,40]

An additional oscillation of 0.19(1) ps appears in the dyad (SI, sec. 3c). The coherent oscillation detected in [Fe-BL-Co] (Fig. 4a) is a combination of signals observed in the photosensitizer and the additional oscillation ($T_{1/2}$=0.19 ps) related to the coordination of cobaloxime. The statistical significance of the superposition could be proven (sec. 3). Calculated excited state potential energy surfaces show that the oscillations appear along the Fe-N bonds with the equilibrium at 2.05 Å ($^3$MLCT*/$^3$MC crossing, see Fig. 5a). TDDFT results indicate several vibrational frequencies in the range around 175 cm$^{-1}$, exhibiting a collective twisting motion of the bridging ligand, accompanied by a rotational distortion and slight stretching of the Fe-N bond. Raman spectra also exhibit intense bands for [Fe-Co-BL] in 175-225 cm$^{-1}$ range, present neither in [Fe-BL], nor in cobaloxime (SI, sec. 3c). While the 0.26 ps half-period can be associated with the spin state transition due to the $^3$MLCT*/$^3$MC crossing, the 0.19 ps oscillation is likely due to the rotation of the cobaloxime moiety around the Fe-Co axis. This motion could affect the charge transfer due to the rotation of the pyridine ring, and modulation of the π* orbitals overlap.

The excited state landscape in the [Fe-BL-Co] dyad can be substantiated with these results. Femtosecond XES study on [Fe(bmip)$_2$]$^{2+}$ showed excited state branching, in which a vibrational wavepacket nearly identical to the one in [Fe-BL-Co] is observed.[16] A $^3$MC is partially populated from the vibrational excited $^3$MLCT* state. Since this wavepacket motion is associated with the MC state,[19] it is not visible in optical TAS measurements. With the minimal spectral difference between $^1$MLCT/$^3$MLCT states both in TAS and XES, the shortest time constant of $\tau_{1,Fe/FeCo}$ in [Fe-BL] and [Fe-BL-Co] is associated with a transition from the $^3$MLCT* to $^3$MC state. The longest time $\tau_{2,Fe/FeCo}$ reflects the $^3$MLCT→$^3$MC pathway. The remaining $\tau_{3,Fe/FeCo}$ is assigned to the $^3$MC→GS recovery.[7,11,17,41,42]

**Population analysis** Kinetic modelling can facilitate the interpretation of the obtained time constants by testing different reaction models. Details of the approach can be found in the supplementary information. At the Fe center in [Fe-BL] and [Fe-BL-Co], an additional time constant of 0.22(7) ps is obtained. According to our TDDFT calculations, this can be related to the $^1$MLCT→$^3$MLCT transition after a population of the first excited $^1$MLCT state (SI, sec. 3e). This short time constant includes IC and ISC.[11,16–18,40] It is also in good agreement with the 350 fs component obtained via TA. According to the proposed reaction scheme, the $^3$MLCT* state, which is populated during the $^1$MLCT→$^3$MLCT*→$^3$MLCT decay, branches into a α ($^3$MLCT*→$^3$MLCT→$^3$MC) and β channel ($^3$MLCT*→$^3$MC), with contributions of 79(5) % and 21(5) %, respectively, as shown in Fig. 5a. In [Fe-BL], the branching ratio is 83 % to 17 %, respectively (SI, sec. 3f). The observed wavepacket oscillations originate from the β pathway.[40]

Most importantly, an additional deactivation channel, originating from the $^3$MLCT state in the form of an M'MCT electron transfer in the [Fe-BL-Co] dyad, is resulting from the kinetic fitting as well. This transfer is clearly visible when the $^3$MLCT population of the pure photosensitizer [Fe-BL] (SI, sec. 3f) and the dyad [Fe-BL-Co] (Fig. 5c) in the short time window is compared. In the former, the rise of the $^3$MLCT population is initially damped, while in the latter, the population of $^3$MLCT rises. The obtained value of the CT rate is very consistent with the magnitude of differences observed for the excited state kinetics at the Co center in [Fe-BL-Co] and cobaloxime (*cf.* Fig. 3).

A two-state model with subsequent decay is used for the cobaloxime (SI, sec. 3f, Fig. S3.9-3.10), consisting of the LMCT state directly populated upon 400 nm excitation and decaying to a lower-level state within 2.78(3) ps. For [Fe-BL-Co] (Fig.5a), an additional electron transfer-acceptor state (M'MCT) is compulsory from the experimental results. The M'MCT decays in 0.25 ps,[21] parallel to the directly excited LMCT decay. The amplitude ratio between the direct excitation and CT transfer yield is 43.0 % to 57.0 %, close to the value obtained via cross-section analysis (SI, sec. 3b, and sec. 3f) and in line with TDDFT results. The kinetic fitting required an additional lowest excited state of this direct decay path, which is of unknown nature so far. However, an MC character is most likely.[43] The lifetime of this state is estimated to be around 23-30 ps based on the fit results for the pure cobaloxime. Data quality for the dyad prevent accurate fitting of this contribution to the fluorescence signal in [Fe-BL-Co].

Fig. 5a summarizes the results and conclusions from the observed time constants, literature,[11,16,17,30,42] and TDDFT potential energy surfaces calculations along two reaction coordinates (Fe – N bite angle and distances). The population analysis for [Fe-BL-Co] resulting from kinetic modelling is shown in Fig. 5b-c (SI, sec. 3f and the corresponding diagram for [Fe-BL]).

## CONCLUSIONS

Photoactive base metal dyads appear as promising alternative, as compared to precious metals, for inexpensive and sustainable molecular assemblies capable of direct harvesting of light and photocatalytic hydrogen production. This still heavily depends on the rational improvement of their performance, which involves the interplay between their molecular design and photocatalytic properties. Our study shows the tremendous potential of ultrafast 2C-XES for direct characterization of photoinduced CT processes exemplified by the case of a noble metal free dyad [Fe-BL-Co] used in hydrogen production. In combination with ultrafast optical spectroscopy, TDDFT and CASSCF/NEVPT2 calculations and excited state modelling, a CT from the Fe$^{II}$ photosensitizer to the cobaloxime catalyst could be proven. It contributes as a M'MCT state of 0.25 ps lifetime to the very complex excited state landscape. In addition, we can distinguish the direct excitation into an LMCT state of Co, which accompanies the CT process between both metals. The unequivocal determination and visualization of the ultrafast CT is only possible by the intrinsic temporal self-calibration of the Fe and Co Kα signals in the 2C-XES experiment.

With the achieved results a multitude of strategies to improve the photocatalytic activity of such base metal dyads can be deduced. It is common knowledge that the lifetime of the $^3$MLCT as the first charge separated state needs to be increased for iron photosensitizers to be active. However, from Fig. 5 it is immediately clear, that this is even more important here. A decreased $^3$MLCT energy would reduce the contribution of the $^3$MLCT→$^3$MC decay channel, potentially in favour of the population of the M'MCT state. Another way of decreasing non-CT decay channels would be a reduction of the $^3$MLCT*→$^3$MC contribution. Since this pathway is connected to the nuclear wavepacket, the associated vibrational motions might play a crucial role. Further restriction of Fe-N oscillations, either *via* replacing N with C atom or construction of a more rigid ligand structure could selectively increase the $^3$MC energy. Both Fe-N and Fe-BL-Co motions are involved here according to the presented results, and substitution of the pyridine by a cyclometalated ligand might be a suitable exchange for Fe-N.

The presented results thus offer a first step towards a rational design of base metal dyads for photocatalytic proton reduction reactions by direct observation and quantification of CT process in functional bimetallic photosensitizer-catalyst assembly by 2C-XES.

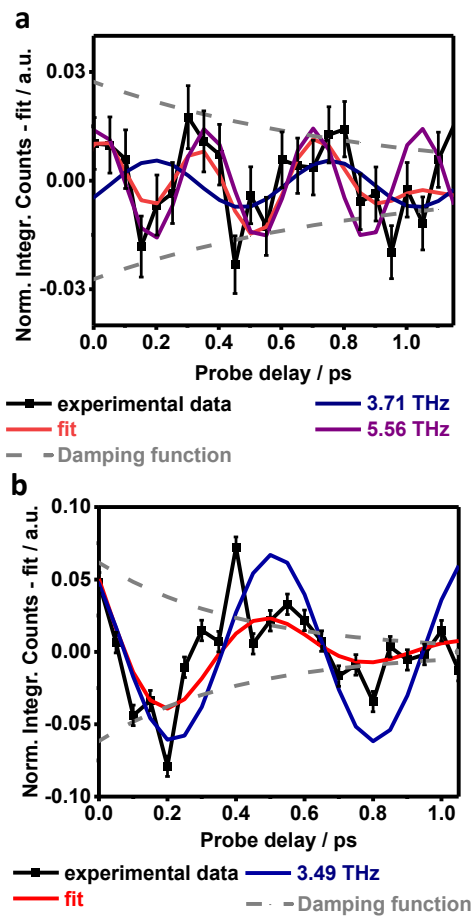

**Fig. 4.** Coherent nuclear wavepacket signals (black), fitted oscillatory functions (red) part, damping (grey) for: **a)** Fe part of [Fe-BL-Co], where additionally a non-damped parts are visible (blue, purple); **b)** same for [Fe-BL].

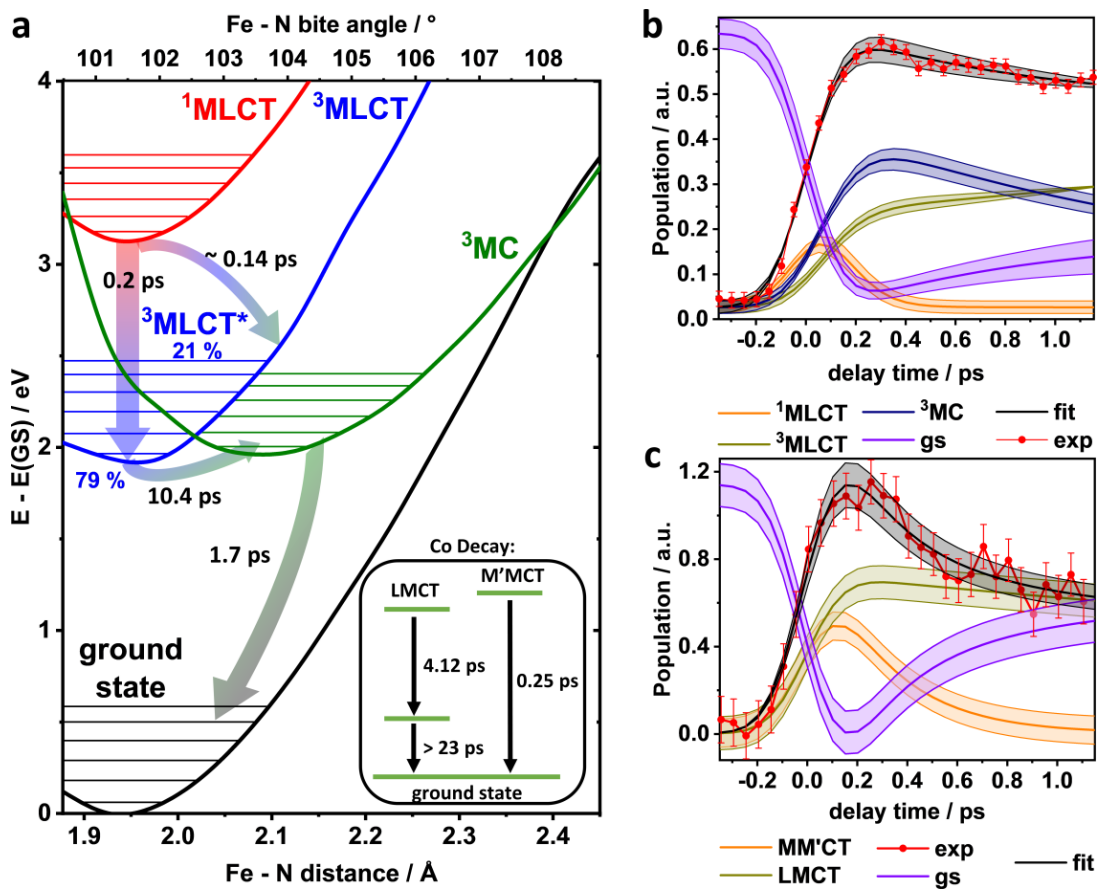

**Fig. 5. (a)** Ground and excited state potential energy surfaces along the Fe-N distance (bottom x-axis) and the Fe-N bite angle (top x-axis). Insert: state diagram for Co. State contributions at the **(b)** Fe in [Fe-BL-Co] **(c)** Co and fs-XES signal.

## METHODS

### UV-Vis spectroscopy

The investigated complexes were dissolved in acetonitrile (spectroscopic-grade, $2.5 \cdot 10^{-4}$ mol/L). UV-Vis spectra were measured in 0.1 cm quartz cuvettes on a Lambda 465 spectrophotometer from PerkinElmer (Waltham, Massachusetts, USA). Cobaloxime ($1 \cdot 10^{-5}$ mol/L) was measured with a Lambda 45 double-beam UV spectrophotometer from Perkin Elmer (Waltham, Massachusetts, USA).

### TAS spectroscopy

The experimental setup was described elsewhere.[5,44] Femtosecond transient absorption dynamic studies of [Fe-BL] and [Fe-BL-Co] were conducted using modified commercial Helios spectrometer (Ultrafast Systems, Sarasota, Florida, USA) with the IRF value of 120 fs. TAS spectra were recorded for the 400 nm excitation in 60 ps temporal range. The laser pulse energy was 2 µJ. Concentrations were chosen to be identical to the time resolved X-ray experiments (10 mM, MeCN), which caused high absorbance of the solutions. Therefore, optimized signal transmission was ensured by a 0.12 µm flow cell with $CaF_2$ windows. The use of a micro annular gear pump (~1 ml/s flow) guaranteed the excitation of fresh solution per laser pulse and reduction of sample degradation. Subtraction of solvent response from each data set eliminated the solvent contribution in the TAS data.

### Transient X-ray emission spectroscopy

Simultaneous emission of Fe and Co Kα was measured with 120 fs time resolution at the FXE instrument at SASE1 branch of EuXFEL, Schenefeld, Germany (Fig. S3.1).[35] The [Fe-BL-Co] dyad in 10 mM solution of acetonitrile (MeCN) was measured in a cylindrical liquid jet (200 µm) and sample recirculation was provided by HPLC pump. Sample was excited by 400 nm optical laser with power in the range of 5.5 - 15 µJ/pulse and 50 fs pulse length (FWHM = 83 µm and 34 µm for horizontal and vertical directions, respectively) which translates to ~55 % of excitation rate. Electronic configuration in ground and excited states were probed by the SASE X-Ray beam with a central energy of 9.3 keV with 125 bunches per pulse train at 0.564 MHz intra-train repetition rate (beam size FWHM = 20 µm, pulse duration 100 fs, ~$10^{12}$ photons/pulse). The X-Ray beam was operating at the standard EuXFEL mode of 10 Hz repetition rate per train and the optical laser was at 5 Hz, meaning alternating pumped/unpumped trains. The beams were crossed with angle of c.a. 20°. Subsequent fluorescence emission was collected using wavelength-dispersive 16-crystal von Hamos XES spectrometer (Fe Kα and Co Kα with Ge(440) and Si(531) analyzer crystal reflections at 75.4° and 77°, respectively) and a 2D charge integrating gain-switching Jungfrau 1M detector with matrix of 1024 x 1024 pixels and repetition rate of 10 Hz. The timing jitter between X-Ray and optical pulses was ~70 fs FWHM. Signal was integrated over 60 s (500 trains) per time point. For different delay time windows, a set of data was acquired with specified temporal step size: for -5ps -15 ps it was 1 ps while for 1.2 ps – 3.3 ps and -1.0 ps – 1.5 ps it was 150 fs. For single delay time measurements, signal was collected for 60 s. For each measurement number of repetitions was set individually to provide good S/N ratio. As a reference also the catalyst cobaloxime and the photosensitizer [Fe-BL] were measured separately in the same experimental conditions and concentrations. Due to limited solubility, cobaloxime was measured at 5 mM.

Quantum chemical calculations

Unless otherwise stated, all calculations were carried out with the ORCA 5.0.1 quantum chemistry package.[45] Throughout we have used the Alrich's def2-TZVP[46] basis set, and employed the Split-RI-J method and chain of spheres (RIJCOSX) approximation to accelerate the calculation of the exchange and Coulomb terms, together with the def2/C and def2/J auxiliary bases.[47] Spin-orbit coupling corrections were introduced using the spin-orbit mean field method.[48] Solvation of the compounds was included via SMD[49] (MeCN) and dispersion correction was introduced via DFT-D3 with the Becke-Johnson damping scheme (D3BJ).[50,51]

Unconstrained DFT optimizations of the investigated complexes were done with the PBEh-3c method.[52–54] The UV-Vis spectra of [Fe-BL] and [Fe-BL-Co] were calculated using the hybrid meta-GGA functional TPSSh[55], employing the Time-dependent DFT (TDDFT) and the Tamm-Dancoff approximation. The adequacy of the method was justified by our benchmark study on the photosensitizer against CASSCF/NEVPT2 (SI, sec. 1a). The singlet energy transitions (60 states) have been subjected to Gaussian broadening with a width of 0.2 eV (full width at half-height) before converting to the nm scale and compared to the experimental UV-Vis spectra of the investigated complexes (*cf*. SI, sec. 1b-c, Fig. S1.3 and S1.4). Donor and acceptor orbitals of selected transitions and their spatial distribution were visualized using Avogadro (*cf*. Table S1.1 and S1.2). Singlet and triplet excited state potential energy surfaces were computed staring from the optimized ground state geometry, by discretizing a geometric pathway that involves a simultaneous stretching of the Fe-N distances at steps 0.05 Å, and the Fe-N bite angles at steps of 0.7 degrees. At each point along this pathway, the 60 lowest lying singlet and triplet states were computed (*i.e.* 120 states in total), again using the aforementioned computational setup. In order to identify the nature of any given excited state, whether it is a $^1$MLCT, $^3$MLCT, or $^3$MC, we have resorted to the Mulliken population analysis coupled with an electron-hole analysis.[56,57] Because our TDDFT calculations are based on the singlet ground state as the reference state, the spin populations of all atoms are zero by symmetry. Instead of relying on spin densities, we identify a $^1$MLCT/$^3$MLCT as a singlet/triplet excited state where the total Mulliken population of the Fe atom is decreased by one electron, and that of the ligand atoms is increased by one electron. A $^3$MC state is a triplet excited state where both the hole and the electron are localized on the Fe atom, corresponding to an electron transfer from the occupied $d_{xy}/d_{yz}/d_{xz}$ orbitals to the virtual $d_{x2-y2}/d_{z2}$ orbitals. In all cases, only excited states that lie below the initially excited $^1$MLCT were considered. Fig. S1.5 depicts an example of this analysis. The geometry of the identified $^3$MC state was optimized and its vibrational normal modes were computed in Gaussian 16 with the def2-SVP basis set.[58]

## ASSOCIATED CONTENT

Any methods, additional references, Nature Research reporting summaries, source data, extended data, supplementary information, acknowledgements, peer review information; details of author contributions and competing interests; and statements of data and code availability are available at:

## ACKNOWLEDGEMENTS

The authors gratefully acknowledge European XFEL in Schenefeld, Germany, for provision of X-ray free-electron laser beamtime at FXE and would like to thank the instrument group and facility staff for their expert assistance. M.B. acknowledges funding by the German DFG in frame of priority program SPP 2102 (Grant number BA 4467/7-1) and the German BMBF (Grant numbers 05K19PP1 and 05K18PPA). W.G. acknowledges partial funding from Narodowe Centrum Nauki through SONATA BIS 6 grant (2016/22/E/ST4/00543). W.G. further acknowledges funding from Spanish MIU through "*Ayudas Beatriz Galindo*" (BEAGAL18/00092), Comunidad de Madrid and Universidad Autónoma de Madrid through Proyecto de I+D para Investigadores del Programa Beatriz Galindo (SI2/PBG/2020-00003), Spanish MICIU through Proyecto de I+D+i 2019 (PID2019-108678GB-I00) and IMDEA-Nanociencia through Severo Ochoa Programme for Centres of Excellence in R&D (MINECO, CEX2020-001039-S). M.H.-G. acknowledges grants by Fonds der Chemischen Industrie and Studienstiftund des deutschen Volkes. Generous grants of computer time at the Paderborner Center for Parallel Computing PC$^2$ is gratefully acknowledged.

## AUTHOR CONTRIBUTIONS

Conceptualization: M.N., M.H.-G., M.B. ; Data curation: M.N., M.H.-G., H.E., J.K.; Formal analysis: M.N., M.H.-G., H.E., J.K, A.K.; Funding acquisition: M.B.; Investigation: M.N., M.H.-G., H.E., J.K., A.K., N.L., D.K., F.L., M.Bv., N.P., P.Z., K.K., A.F.-R., W.G., M.B.; Methodology: M.N., M.H.-G., W.G., T.K., M.B.; Project administration: M.B.; Resources: M.H.-G., M.B.; Software: M.N., M.Bv., H.E.; Supervision: W.G., T.K., M.B.; Validation: M.N., H.E., W.G., M.B.; Visualization: M.N., M.H.-G., H.E., J.K., W.G., M.B.; Writing – original draft: M.N., M.H.-G., H.E., J.K., W.G.; Writing – review & editing: M.N., M.H.-G., H.E., J.K., A.K., N.L., D.K., F.L., T.-K.C., M.Bv., N.P., P.Z., K.K., A.F.-R., W.G.,T.K., M.B..

## DATA AVAILABILITY STATEMENT

The datasets generated during and/or analysed during the current study are available from the corresponding author on reasonable request.

## COMPETING INTERESTS

The authors declare no competing interests.

## ADDITIONAL INFORMATION

**Extended data** is available for this paper at

**Supplementary information** is available for this paper at

**Correspondence and requests for materials** should be addressed to M.B.

**Reprints and permissions information** is available at www.nature.com/reprints

# Ultrafast two-colour X-ray emission spectroscopy reveals excited state landscape in a base metal dyad

M. Nowakowski[1†], M. Huber-Gedert[1†], H. Elgabarty[1], J. Kubicki[2], A. Kertem[2], N. Lindner[2], D. Khakhulin,[3] F. Lima[3], T.-K. Choi[3,4], M. Biednov[3], N. Piergies[5], P. Zalden[3], K. Kubicek[3], A. Rodriguez-Fernandez[3], M. Alaraby Salem[1], T. Kühne[1], W. Gawelda[2,6,7], M. Bauer[1*]

[1] Chemistry Department and Center for Sustainable Systems Design (CSSD), Faculty of Science, Paderborn University, Warburger Straße 100, 33098 Paderborn, Germany

[2] Faculty of Physics, Adam Mickiewicz University, Uniwersytetu Poznańskiego 2, Poznań, 61-614, Poland

[3] European X-Ray Free-Electron Laser Facility GmbH, Holzkoppel 4, 22869 Schenefeld, Germany

[4] PAL-XFEL, Jigok-ro 127-80, 37673 Pohang, Republic of Korea

[5] Institute of Nuclear Physics Polish Academy of Sciences, Kraków, 31-342, Poland.

[6] Departamento de Química, Universidad Autónoma de Madrid, Campus Cantoblanco, 28047 Madrid, Spain

[7] IMDEA Nanociencia, Calle Faraday 9, 28049 Madrid, Spain

# Supplementary Information

Table of contents:

1) Quantum chemical calculations:
   a) Benchmarking the TDDFT UV-Vis spectra of [Fe-BL]
   b) Computed and experimental UV-Vis spectrum of [Fe-BL]
   c) Computed and experimental UV-Vis spectrum of [Fe-BL-Co]
   d) Mulliken population-based electron-hole analysis of excited states
   e) Decomposing the UV-Vis spectrum of [Fe-BL] in terms of charge-transfer components
   f) Decomposing the UV-Vis spectrum of [Fe-BL-Co] in terms of charge-transfer components
2) TA experimental analysis
3) XES data analysis:
   a) Fluorescence fitting procedure and results
   b) A direct and non-direct contribution to Kα XES
   c) Wavepacket analysis
   d) Co Kα1 kinetic signals for -5-15 ps time window
   e) d'-d interactions in Co
   f) Kinetic model and population analysis results

# 1. Quantum chemical calculations

### a) Benchmarking the TDDFT UV-Vis spectra

In order to understand the optical spectrum and the nature of the optical excitation process, we have resorted to quantum chemical calculations using time-dependent density functional theory (TDDFT). It is generally true that the study of transition metal complexes is challenging because of dynamic correlation effects, system size, state degeneracies or near-degeneracies, and relativistic effects on top of the typically large system sizes. In particular, TDDFT is known to have difficulties with systems having charge-transfer states, and with extended $\pi$-systems[1,2], both are features of the dyad. However, despite these well-known issues, TDDFT has nevertheless been successfully applied to study such systems, including $d^6$ transition metal complexes.[3] These known issues mean however, that one should not blindly trust TDDFT results without scrutiny.

To this end, we have benchmarked TDDFT UV-Vis electronic spectra, using both the hybrid-GGA B3LYP functional and the hybrid-meta-GGA TPSSh functional, against CASSCF-NEVPT2 spectra. While CASSCF/NEVPT2 is known to reliably yield reasonable accuracy,[4] the dyad molecule is too large for the method. The CASSCF/NEVPT2 method explicitly takes account of both static and dynamic correlation effects and is known to provide highly accurate spectra.[5] In order to keep the size of the active space manageable we have done the benchmarking against the photosensitizer without the cobaloxime moiety. As explained later, the active space required to accurately compute the electronic spectrum of the photosensitizer included 14 electrons in 13 active orbitals. A CASSCF/NEVPT2 of the dyad, including all the 12 $d$-electrons together with the interacting ligand electrons was computationally unfeasible due to the large number of occupied orbitals in the active space.

Our benchmark calculations show that the TPSSh functional yields qualitatively correct result and accurately reproduces the spectrum with a slight tendency to over-estimate the frequency of the peaks, in particular the lowest-frequency peak. To obtained better comparison of calculated spectra with experimental ones calculated spectra are broadened by convolution with a Gaussian function with a width of 0.2 eV (full width at half-height), before converting the scale to nm.

**Starting orbitals for CASSCF**

The starting orbitals for the CASSCF calculation were taken from the TPSSh ground-state Kohn-Sham orbitals at the equilibrium geometry. The TPSSh ground state has the close-lying (within 0.3 eV) Fe $d_{xy}$, $d_{yz}$, and $d_{xz}$ orbitals as the three occupied frontier orbitals, these were naturally included in the active space. The $d_{z2}$ and the $d_{x2-y2}$ orbitals were found to be strongly mixed with ligand orbitals, consistent with the strongly σ-donating heterocyclic carbene ligand. Both the occupied (bonding) and unoccupied (antibonding) orbitals involving Fe $d_{x2-y2}$ and $d_{z2}$ were included in the active space. In addition to the full set of (ligand-mixed) Fe $d$-orbitals, the two highest lying occupied π-bonding orbitals were included in the active space, together with the four lowest unoccupied molecular orbitals (LUMO to LUMO+3). The LUMO is a π* orbital extending over the bipyridine moiety, while the other three orbitals are all π* orbitals extending on the CNC moieties. Thus, the final active space included 14 electrons in seven occupied orbitals and six virtual orbitals.

It is worth mentioning here that the B3LYP Kohn-Sham orbitals were identical in character to the TPSSh orbitals, in agreement with the benchmark results that we discuss below.

**Comparison of TDDFT UV-Vis spectra to CASSCF(14,13)/NEVPT2**

The obtained spectra, which are depicted in Fig. S1.1, show several interesting features. The CASSCF/NEVPT2 spectrum closely follows the experimental one, with two major peaks at 456.2 and 389.0 nm. We believe that the major source of the shift from the experimental spectrum is the implicit solvation model. Between these two major absorption peaks, there is a weak absorption peak at 422.1 nm (~10% of the oscillator strength of the strong peaks). The TDDFT spectra, although blue-shifted, still provide qualitatively correct results, except for the wrong trend in the peak intensities, with the low-frequency peak having a lower amplitude than the high-frequency one. The TPSSh functional is clearly performing better than B3LYP, with the TPSSh peaks appearing at 446.0, 415.2, and 395.9 nm, compared to 411.9, 387.1, and 365.1 nm for B3LYP. The accuracy of TDDFT transition frequencies, which we find here, is consistent with the expected accuracy range of the method, typically within 0.1-0.5 eV.[5,6] The lack of any peaks below 300 nm in the CASSCF/NEVPT2 spectrum is because here we have only calculated the twelve lowest-lying singlet states (compared to 60 states in TDDFT).

Rather than the exact positions of the peaks, more important to our benchmark is the nature of the underlying states. Here, we find very consistent behavior between TDDFT (both functionals) and CASSCF. Both methods agree that the main transitions bear predominantly the MLCT character and originate from the three frontier orbitals to the virtual orbitals in the range LUMO – LUMO+3. The low-frequency peak is consistently the transition Fe $d_{yz}$ → LUMO with contribution from the transition Fe $d_{xy}$ → LUMO+2 orbital. Also, all the methods show that the higher frequency peak is mainly $d_{xy}$ → LUMO+3 with a minor contribution from $d_{xy}$ → LUMO+1, and that the weak intermediate frequency peak is a transition from the three frontier orbitals to the three virtual orbitals LUMO+1 – LUMO+3 (for details see Table S1.1).

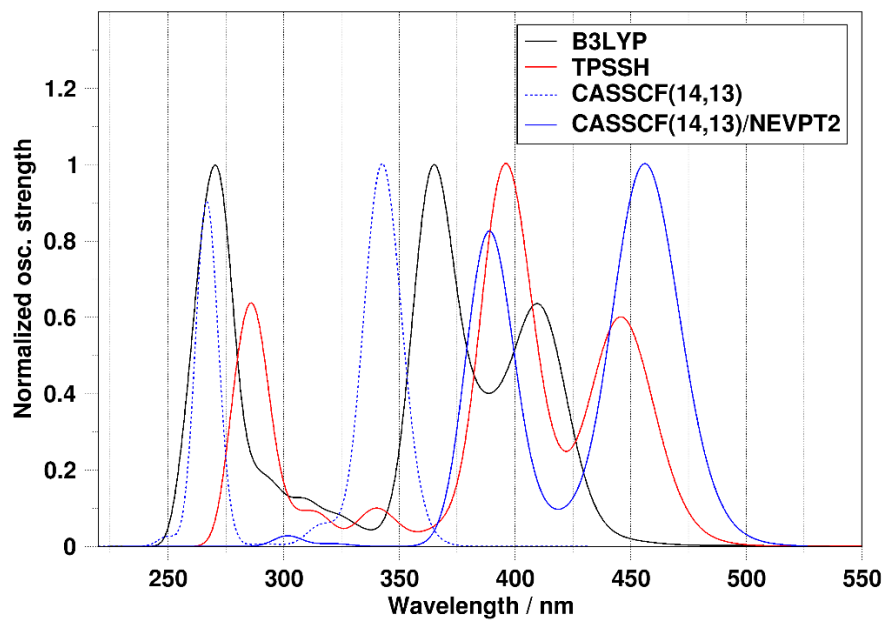

**Fig. S1.1.** UV-Vis absorption spectrum of the photosensitizer in implicit acetonitrile solvation. Black: TDDFT with B3LYP, red: TDDFT with TPSSh, dotted blue: CASSCF(14,13), blue: CASSCF(14,13)/NEVPT2. All the spectra are broadened by convolution with a Gaussian function with a width of 0.2 eV (full width at half-height), before converting the scale to nm.

**Influence of spin-orbit coupling**

In computing all the TDDFT spectra, we have included corrections due to spin-orbit coupling. It is worth noting however, that this turned out to have very little influence on peak positions, with typical shifts of less than 1 nm. As an example, Fig. S1.2 shows the influence of spin-orbit coupling on the UV-Vis spectrum of the photosensitizer, as obtained with B3LYP/TDDFT.

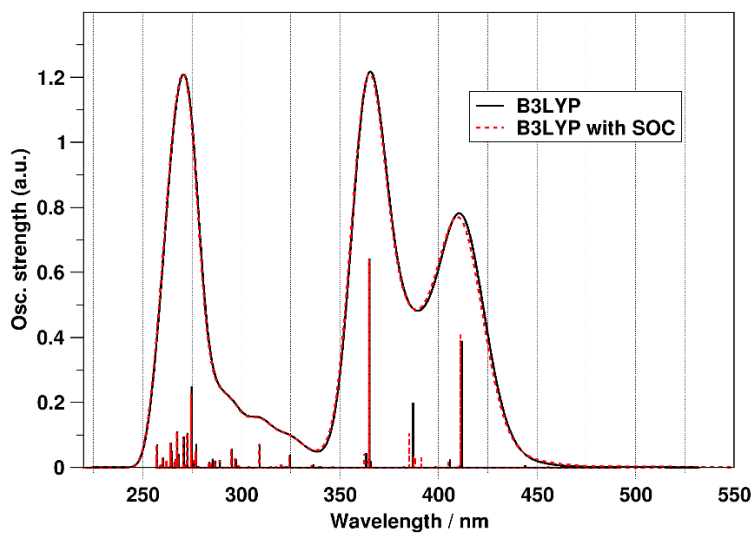

**Fig. S1.2.** Influence of spin-orbit coupling (SOC) on the UV-Vis spectrum of the photosensitizer.

**b) Computed and experimental UV-Vis spectrum of [Fe-BL]**

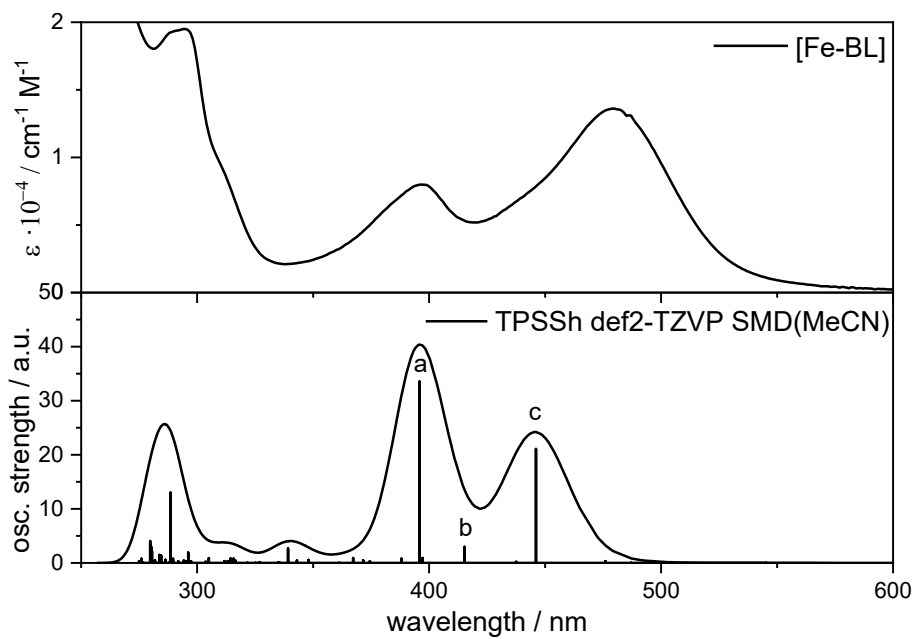

**Fig. S1.3** Experimental UV-Vis spectrum of [Fe-BL] in MeCN and time-dependent TDDFT spectrum with TPSSh.

**Table S1.1** Computed dominant singlet vertical excitations a-c of [Fe-BL]. Donor and acceptor orbitals are listed together with their contribution to the transition. The main character of the transition is indicated.

| Transition (state) | Donor | Acceptor | Contribution | Character |
|---|---|---|---|---|
| a (8) 395.9 nm | HOMO (237) 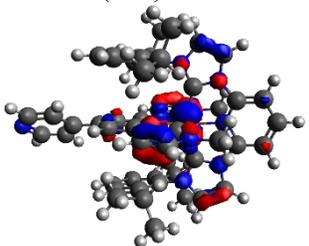 | LUMO+3 (241) 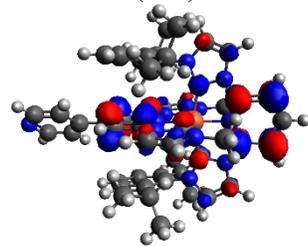 | 0.73 | MLCT |
| | HOMO-2 (235) 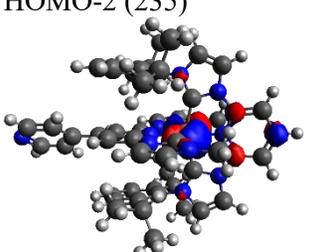 | LUMO+1 (239) 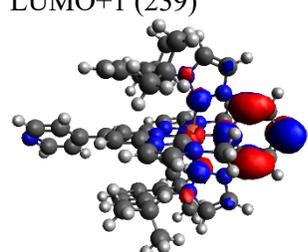 | 0.10 | MLCT |
| b (6) 415.2nm | HOMO-2 (235) 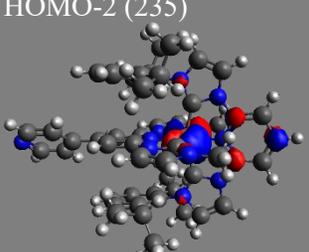 | LUMO+1 (239) 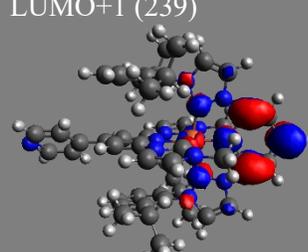 | 0.37 | MLCT |
| | HOMO (237) 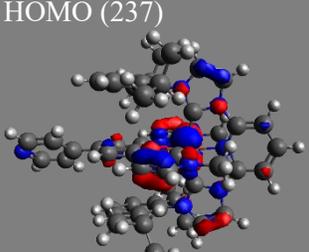 | LUMO+2 (240) 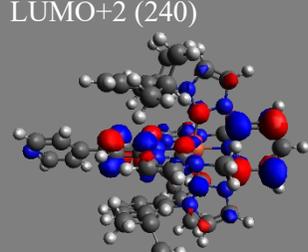 | 0.36 | MLCT |
| | HOMO (237) 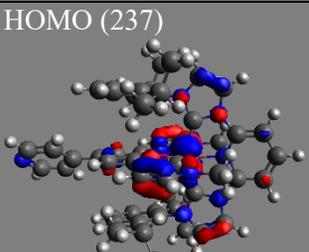 | LUMO+3 (241) 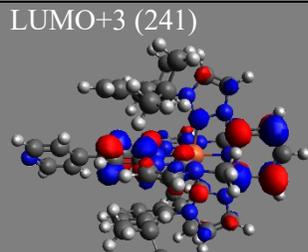 | 0.22 | MLCT |
| c (4) | HOMO-1 (236) 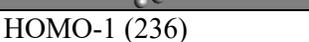 | LUMO (238) 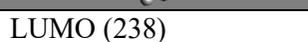 | 0.64 | MLCT |

| 446.0 nm | | | | |
|---|---|---|---|---|
| | HOMO (237) | LUMO+2 (240) | 0.20 | MLCT |
| | HOMO-2 (235) | LUMO (238) | 0.09 | MLCT |

**c) Computed and experimental UV-Vis spectrum of [Fe-BL-Co]**

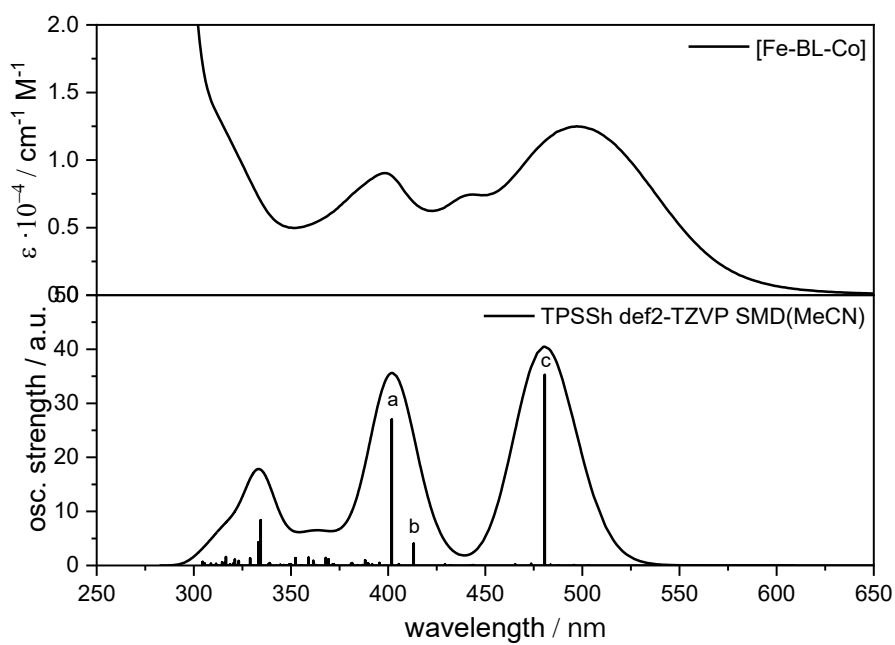

**Fig. S1.4** Experimental UV-Vis spectrum of [Fe-BL-Co] in MeCN and time-dependent TDDFT spectrum with TPSSh.

**Table S1.2** Computed dominant singlet vertical excitations a-c of [Fe-BL-Co]. Donor and acceptor orbitals are listed together with their contribution to the transition. The main character of the transition is indicated.

| Transition (state) | Donor | Acceptor | Contribution | Character |
|---|---|---|---|---|
| a (17) 401.9 nm | HOMO (320) 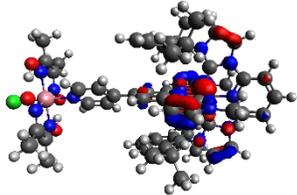 | LUMO+7 (328) 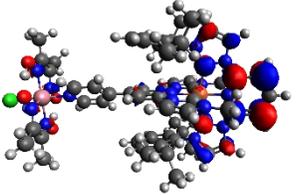 | 0.37 | MLCT |
| | HOMO (320) 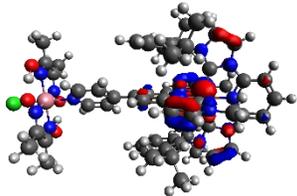 | LUMO+4 (325) 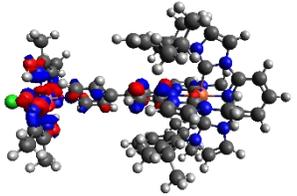 | 0.29 | MLCT/ MMCT |
| | HOMO (320) 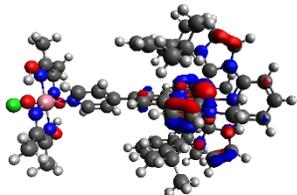 | LUMO+5 (326) 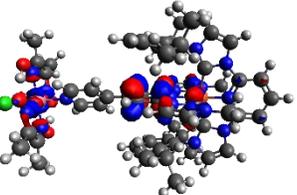 | 0.21 | MLCT/ MMCT |
| b (15) 413.1 nm | HOMO (320) 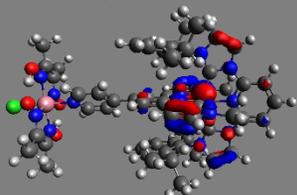 | LUMO+7 (328) 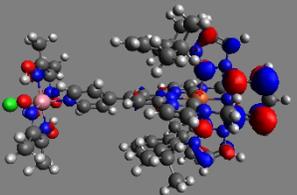 | 0.38 | MLCT |
| | HOMO-2 (318) 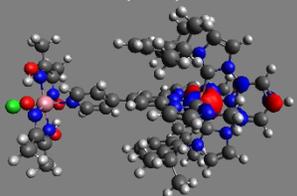 | LUMO+3 (324) 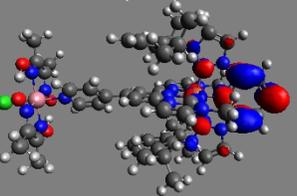 | 0.37 | MLCT |
| | HOMO (320) 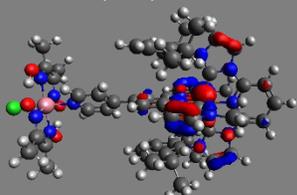 | LUMO+5 (326) 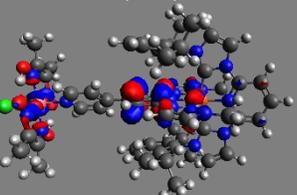 | 0.14 | MLCT/ MMCT |
| | HOMO-1 (319) | LUMO (321) | 0.78 | MLCT |

| c (8) 480.6 nm | | | | |
|---|---|---|---|---|
| | HOMO (320) | LUMO+5 (326) | 0.08 | MLCT/ MMCT |

The TDDFT calculation of the dyad indicates transitions with partial MMCT character for the UV-Vis band around 400nm and 480nm. The charge transfer from iron to cobalt is further analyzed by the charge transfer analysis in section 1f.

### d) Mulliken population-based electron-hole analysis of excited states

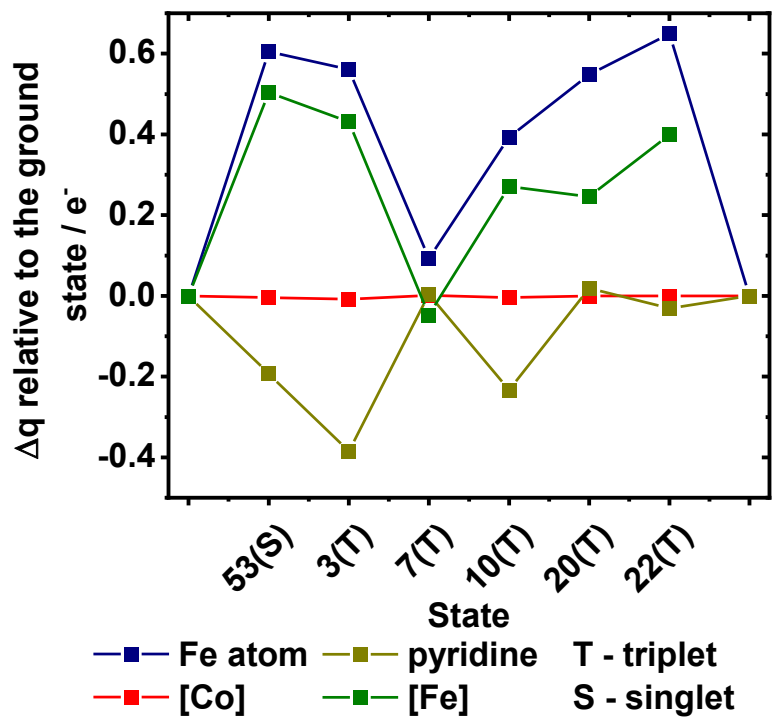

**Fig. S1.5.** Mulliken population analysis of the triplet excited states that show the strongest metal-centered character. State number 53 is the initially populated $^1$MLCT state. "Fe" and "Co" refer to the Mulliken populations of the two metal atoms, "pyridine" is the total charge on the bridge pyridine attached to the Fe, and "Fe-coord" is the octahedral coordination.

In order to identify the nature of any given excited state as obtained from TDDFT, whether it is a $^1$MLCT, $^3$MLCT, or $^3$MC, we have resorted to the Mulliken population analysis coupled with an electron-hole analysis.[7] Because our TDDFT calculations are based on the singlet ground state as the reference state, the spin populations of all atoms are zero by symmetry. Instead of relying on spin densities, we identify a $^1$MLCT/ $^3$MLCT as a singlet/triplet excited state where the total Mulliken population of the Fe atom is decreased by one electron, and that of the ligand atoms is increased by one electron. A $^3$MC state is a triplet excited state where both the hole and the electron are localized on the Fe atom, corresponding to an charge transfer from the occupied $d_{xy}/d_{yz}/d_{xz}$ orbitals to the virtual $d_{x2-y2}/d_{z2}$ orbitals. In all cases, only excited states that lie below the initially excited $^1$MLCT were considered.

Fig. S1.5 graphically depicts the outcome of such an analysis on the optimized geometry of the singlet ground state. In this particular case, state 7(T) is readily identified as the lowest lying $^3$MC state. Calculation of the Mulliken population contribution of the Fe atom to the hole and electron redistribution confirms the identity of this state, with the Fe atom contributing 84.8% to the electron hole (*i.e.* the excited electron originates from the Fe), and with 68% of the redistributed electron density concomitantly residing on the Fe atom (*i.e.* the excited electron resides on the Fe).

### e) Decomposing the UV-Vis spectrum of [Fe-BL] in terms of charge-transfer components

This qualitative characterization of the MLCT charge-transfer nature of the main transitions in the spectrum like the one in Fig. S1.1 can be put into more quantitative terms using a hole-electron analysis.[7,8] The idea here is to start with the usual expression for the UV-Vis spectrum as obtained via broadening the excitation energies of all excited states:

$$\varepsilon(E) \propto \sum_i f_i\, G(E - E_i^{exc.})$$

Where $E_i^{exc.}$ is an excitation energy, $f_i$ the corresponding oscillator strength, and G(…) denotes convolution with a lineshape function (A Gaussian function in this work). If we now subdivided the molecule into two mutually exclusive fragments A and B (generalization to more fragments is trivial), then the excitation spectrum can be readily decomposed as:

$$\varepsilon(E)_{A,B} \propto \sum_i f_i Q_i^{A,B} G(E - E_i^{exc.})$$

where $Q_i^{A,B}$ is the amount of charge transfer from A to B in excited state $i$ as obtained, in this case, by a Mulliken population analysis. Because the sum of all inter- and intra-fragment charge transfer terms is unity, the partitioning is exact, and the total spectrum is exactly divided into two intra-fragment (A→ A and B→ B) charge redistribution terms and two inter-fragment (A→ B and B→ A) charge transfer terms.

To decompose the photosensitizer spectrum in Fig. S1.6, the structure was subdivided into three fragments: the iron atom (fragment 1), the terminal bipyridine moiety (fragment 3), and the rest of the molecule (fragment 2). Figure S1.6 depicts the decomposed TPSSh/TDDFT spectrum. The decomposed spectrum clearly reveals the nature of all the peaks in the spectrum. For instance, the low-frequency peak involves mainly (50% of the total amplitude) a charge transfer from the iron to the terminal bipyridine, where the LUMO orbital resides, but also includes important 1→ 2 and 1→ 3 charge-transfer contributions. On the other hand, the 1→ 2 charge transfer spectrum is most prominent in the peak close to 400 nm, but also the shoulder due to the contribution of the weak intermediate peak centered at 415.2 nm is also clear.

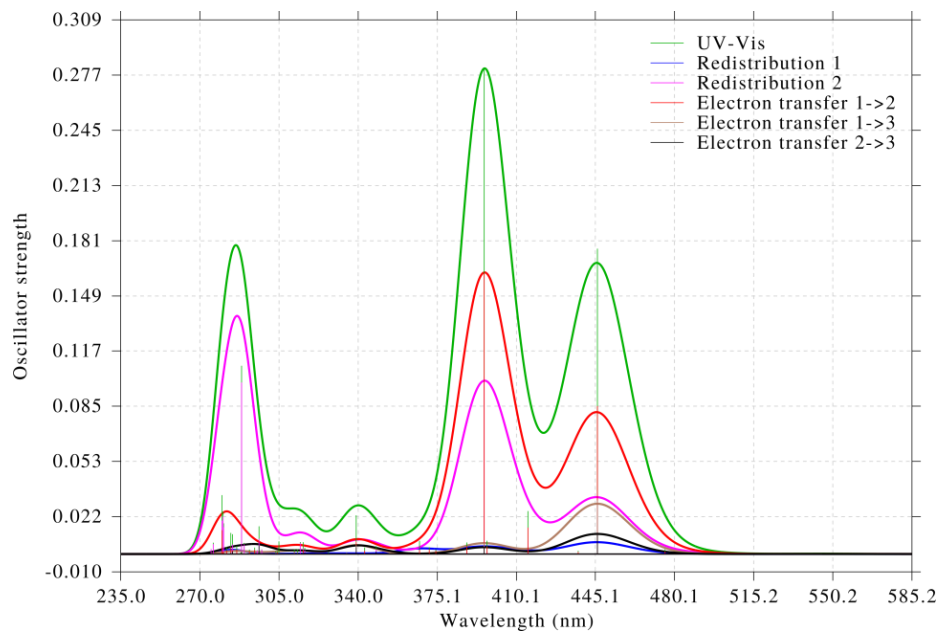

**Fig. S1.6.** Decomposition of total UV-Vis spectrum into intrafragment charge redistribution and interfragment charge transfer contributions. Fragment 1 is the iron atom, fragment 3 is the terminal pyridine moiety, and fragment 2 is the rest of the molecule.

### f) Decomposing the UV-Vis spectrum of [Fe-BL-Co] in terms of charge-transfer components

In an analogic way to the [Fe-BL] case, we decomposed the UV-Vis spectrum of the [Fe-BL-Co]. Fig. S1.7 shows separate parts of the dyad considered in this analysis along with the color code. Table S1.3 presents contributions to the total charge for all considered transitions for each of the molecular parts shown in Fig. S1.7.

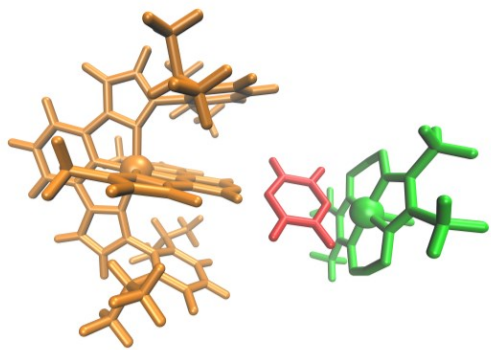

**Fig. S1.7.** Definition of the three fragments used to decompose the UV-Vis spectrum. Orange: fragment 1, red: fragment 2, green: fragment 3

**Table S1.3.** The fractional contribution of each fragment to the hole and the electron in each of the three major transitions in the spectrum.

| Wavelength / nm | oscillator strength | hole(1) | electron(1) | hole(2) | electron(2) | hole(3) | electron(3) |
|---|---|---|---|---|---|---|---|
| 480.6 | 0.2949 | 0.976 | 0.560 | 0.016 | 0.414 | 0.008 | 0.026 |
| 413.1 | 0.0338 | 1.000 | 0.989 | 0.000 | 0.010 | 0.000 | 0.000 |
| 401.9 | 0.2259 | 0.993 | 0.898 | 0.002 | 0.073 | 0.006 | 0.029 |

Fig. S1.8 presents the charge transfer analysis. It reveals that the peak at ~400 nm has the same nature as in the photosensitizer, with a small fraction of Fe → Co charge transfer (shown directly in Fig. 1e). The low-frequency peak corresponds to considerably more charge transfer to the terminal pyridine ring.

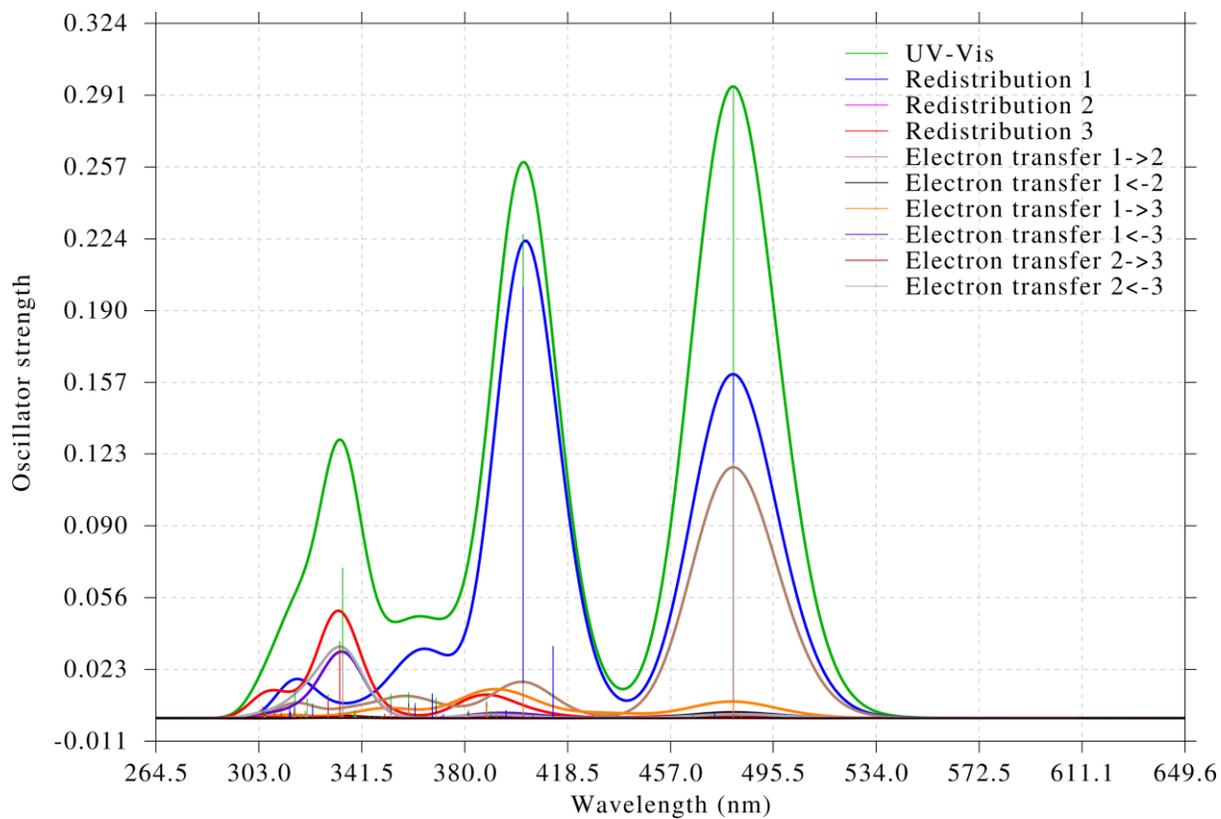

**Fig. S1.8.** Decomposition of total UV-Vis spectrum into intrafragment charge redistribution and interfragment charge transfer contributions. Fragment 1 is the iron atom moiety, fragment 2 is the terminal pyridine moiety, and fragment 2 is the cobalt atom moiety.

## 2. TAS data analysis

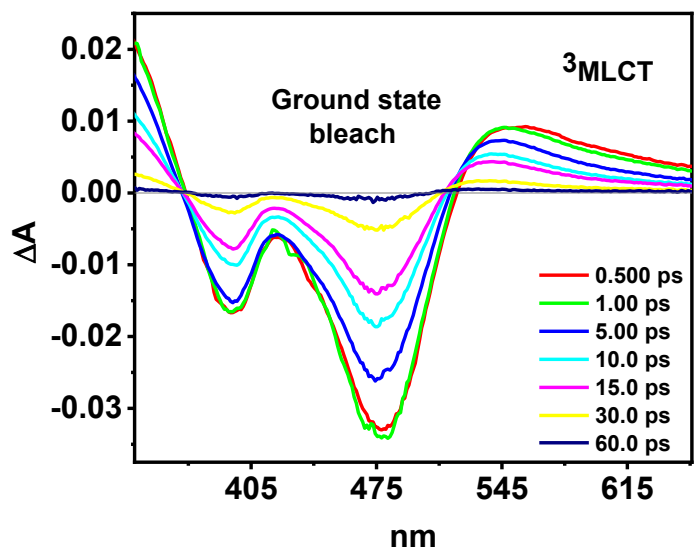

**Fig. S2.1.** TAS spectra recorded upon 400 nm excitation for the [Fe-BL].

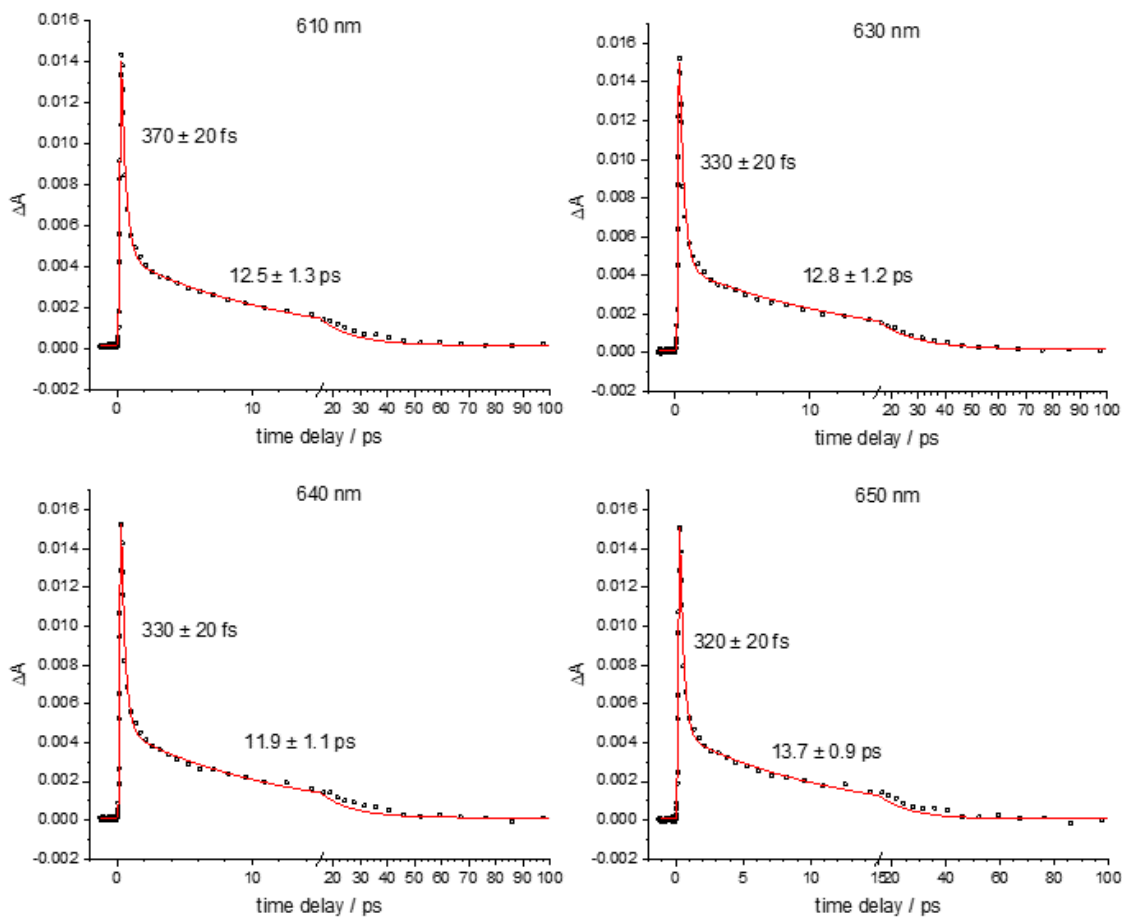

**Fig. S2.2.** Kinetics recorded for [Fe-BL-Co] at 650 nm, 640 nm, 630 nm and 610 nm together with the fitted model (red lines).

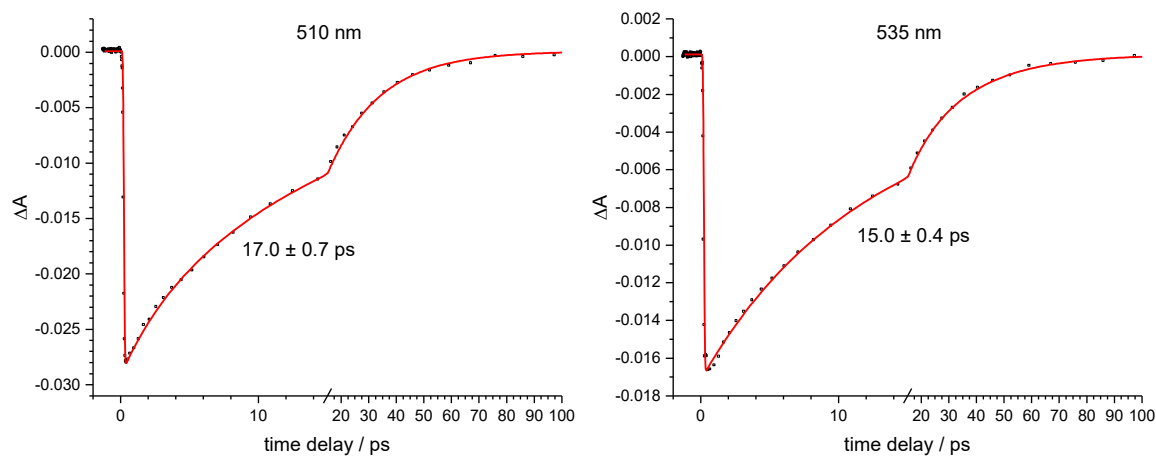

**Fig. S2.3**. Kinetics recorded for [Fe-BL-Co] at 510 nm and 535 nm presenting the temporal evolution of the recovery of the GS together with the fitted model (red lines).

It is commonly accepted that global analysis is applied to reveal a temporal evolution of such complexes but in this case as our model assumes that the ground state (GS) is repopulated by $^3$MC and not the $^3$MLCT (here, see Fig. 5a) we did not use this approach. Therefore, the recovery of the GS is expected to be slower than the decay of the $^3$MLCT state, although both relaxation channels are occurring on the same order of magnitude, i.e., few tens of picoseconds. Kinetics in 510-535 nm spectral range provide that the recovery of the GS takes place with a time constant in the range from 15.0 - 17.0 ps. The $^3$MLCT-related time constant changes a little (12.1 ps - 12.7 ps) depending on the wavelength selected for the strongest GS bleach band (Figure S2.2). It is very likely due to the vibrational cooling of the hot GS.[9] Few picosecond longer recovery of GS than the decay of $^3$MLCT state is consistent with the fact that the lifetime of $^3$MC state is of the order of 2 ps (*vide infra*).[10–14] The same effect can be observed in kinetics extracted for the [Fe-BL] 630 nm, 550 nm, and 480 nm (Fig. S2.4). The comparison with the kinetic data between the photosensitizer and the dyad can be done only qualitatively since upon formation of the bimetallic assembly the 470 nm feature in [Fe-BL] UV-Vis spectrum relaxes to ~500 nm.

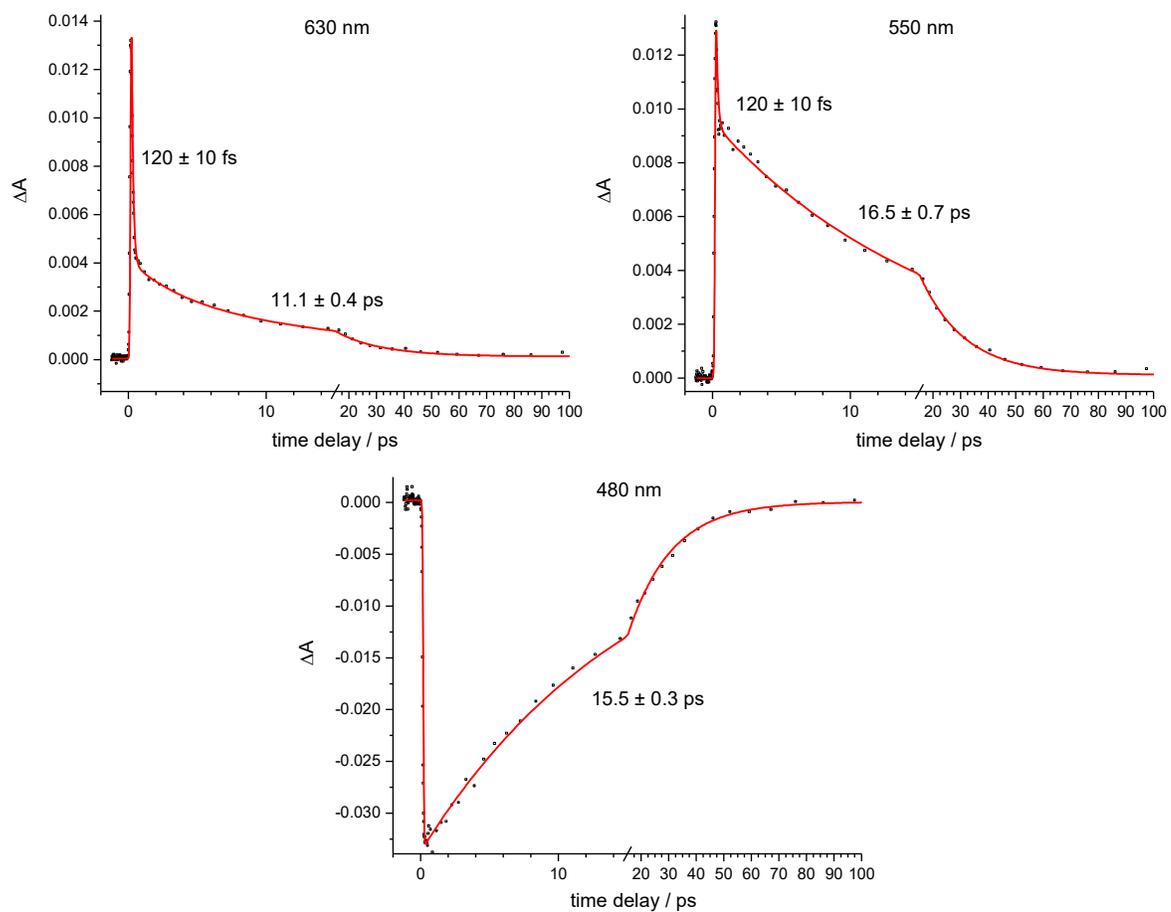

**S2.4.** Kinetics recorded for [Fe-BL] at 630 nm, 550 nm and 480 nm together with the fitted model (red lines).

# 3. XES data analysis

### a) Fluorescence fitting procedure and results

The time-resolved X-ray emission spectroscopy (TR-XES) data was obtained on a setup scheme presented in Fig. S3.1.

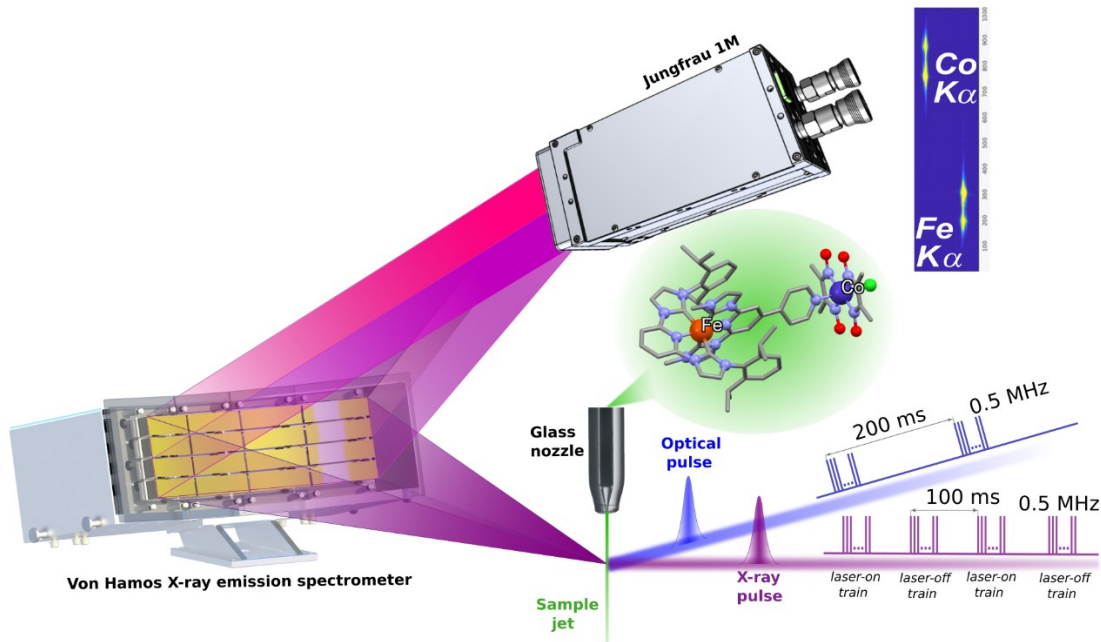

**Fig. S3.1.** Scheme of the experimental setup used at FXE beamline. 50 fs FWHM optical pulses of 400 nm wavelength (blue) were synchronized with 100 fs FWHM X-ray pulses (purple) with the timing jitter of ~70 fs. Fe and Co Kα X-ray fluorescence emitted from the liquid jet (green) sample was analyzed by a 16-crystal array of von Hamos spectrometer and directed to the 2D Jungfrau detector

The energy scale for TR-XES data was obtained on a basis of the ground state XES (gs-XES) measurements performed using von Hamos spectrometer at P64 beamline of Petra-3 synchrotron at DESY (Hamburg). The gs-XES energy calibration was obtained by measuring Fe foil and adjusting the first inflection point in XAS spectrum to 7112 eV. Initial data correction: empty pulse removal and dark correction were conducted on-site, while data reduction and extraction were performed remotely on DESY Maxwell server with the use of self-written Python scripts. A set of data from the experiment was sorted, background reduced, filtered, and normalized to obtain ON/OFF XES spectra in respect to delay time between the optical pump and X-ray probe pulses. From that a series of XES spectra, differential (transient spectra, ΔXES) spectra were calculated, as $\Delta XES(t) = XES_{ON}(t) - XES_{OFF}(t)$, both for Fe and Co Kα lines, examples are in Fig. S3.1. Progressing changes in the ΔXES profile were represented in form of the integral of the selected feature over all delay times. Those kinetics were subsequently fitted with fluorescence rise and $i$–exponential decay functions to give decay rates $\tau_i$ [15]:

$$y = y_0 + \sum_i A_i g_i(t) \quad \text{(S3.a.1)}$$

$$g_i(t) = \frac{1}{2}\left(1 + erf\left(\frac{t-t_0}{\sigma}\frac{C}{2} - \frac{\sigma}{2C\tau_i}\right)\right) e^{\frac{\sigma^2}{C^2\tau_i^2}} e^{-\frac{t-t_0}{\tau_i}} \quad \text{(S3.a.2)}$$

where:

$C = 2\sqrt{ln(16)}$ ;

$A_i$ – amplitude for the $i$–th exponent;

$t_0$ – the time-zero constant value;

$\sigma$ – Gaussian broadening due to IRF function. Due to used setup the IRF was fixed to 0.28 ps;

$i = 1,2,3$ – the degree of exponential function.

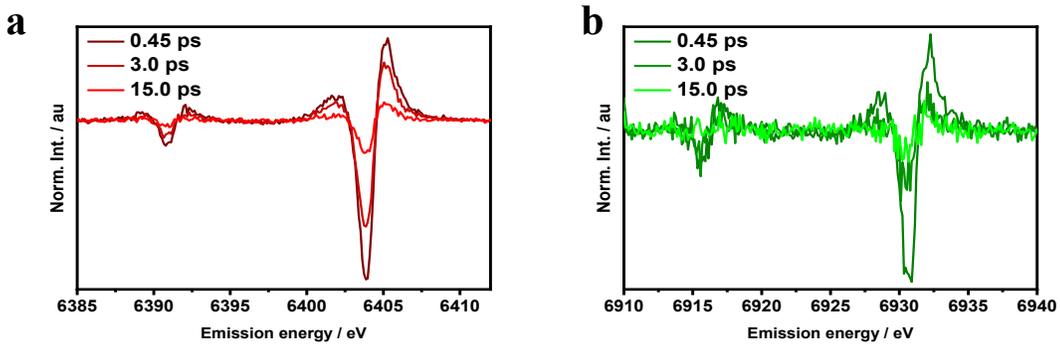

**Fig. S3.2.** K$\alpha_1$ transient XES for: **a)** Fe @ [Fe-BL-Co]; **b)** Co @ [Fe-BL-Co].

The fitting procedure was carried out in two steps. First, the largest dominating contribution was fitted to a kinetic trace in the time window of 20 ps and step size of 1 ps. Second, to fit smaller decay time constants, kinetic traces in time windows of ~2 ps and step size of 50 fs with the largest time constant were fixed. All fitting procedures were performed with a value of FWHM in pump-probe cross-correlation function set to $\sigma = 0.284$ ps. The $\sigma$ value was refined in the post-fitting verification. The summarized fitting results are presented in Table S3.1 and in Figs. S3.2-3.

**Table S3.1. Summary for fluorescence fitting to experimental data.**

| [Fe-BL] 15 ps | | [Fe-BL] 1.2 ps | |
|---|---|---|---|
| $A_1$ | - | $A_1$ | 0,170(26) |
| $\tau_1$ [ps] | - | $\tau_1$ [ps] | 0,245(42) |
| $A_2$ | 0,241(5) | $A_2$ | 0,538(7) |
| $\tau_2$ [ps] | 8,984(273) | $\tau_2$ [ps] | 10,142(2,054) |
| $A_3$ | 0,079(12) | $A_3$ | 0,126(8) |
| $\tau_3$ [ps] | 1,705(348) | $\tau_3$ [ps] | 2,421(640) |
| $t_0$ [ps] | 0,368(31) | $t_0$ [ps] | 0,015(7) |
| $y_0$ | 0,070(2) | $y_0$ | 0,061(6) |
| FWHM[a] [ps] | 0,289(61) | FWHM[a] [ps] | 0,305(21) |

| Fe @ [Fe-BL-Co] 15 ps | | Fe @ [Fe-BL-Co] 1.2 ps | |
|---|---|---|---|
| $A_1$ | - | $A_1$ | 0,109(20) |
| $\tau_1$ [ps] | - | $\tau_1$ [ps] | 0,115(23) |
| $A_2$ | 0,00135(2) | $A_2$ | 0,483(3) |
| $\tau_2$ [ps] | 10,381(242) | $\tau_2$ [ps] | 12,417(1,399) |
| $A_3$ | - | $A_3$ | 0,111(4) |
| $\tau_3$ [ps] | - | $\tau_3$ [ps] | 1,740(182) |
| $t_0$ [ps] | -0,064(12) | $t_0$ [ps] | 0,010(3) |
| $y_0$ | 4,922(56)·$10^{-4}$ | $y_0$ | 0,024(2) |
| FWHM[a] [ps] | 0,284(48) | FWHM[a] [ps] | 0,275(10) |

| Co @ [Fe-BL-Co] 15 ps | | Co @ [Fe-BL-Co] 1.2 ps | |
|---|---|---|---|
| $A_1$ | 3.140(421)·$10^{-4}$ | $A_1$ | 0,00257(13) |
| $\tau_1$ [ps] | 0.25 (fixed) | $\tau_1$ [ps] | 0,242(14) |
| $A_2$ | 1.249(175)·$10^{-4}$ | $A_2$ | - |
| $\tau_2$ [ps] | 4.12(1.39) | $\tau_2$ [ps] | - |
| $A_3$ | 1.381(149)·$10^{-4}$ | $A_3$ | 0,00187(37) |
| $\tau_3$ [ps] | 23.39 (fixed) | $\tau_3$ [ps] | 6,084(1,134) |
| $t_0$ [ps] | -0,116(10) | $t_0$ [ps] | 0,047(7) |
| $y_0$ | 1,049(31)·$10^{-4}$ | $y_0$ | 8,061(282)·$10^{-4}$ |
| FWHM[a] [ps] | 0,284(42) | FWHM[a] [ps] | 0,280(19) |

| Cobaloxime 215 ps | | Cobaloxime 1.2 ps | |
|---|---|---|---|
| $A_1$ | 2.342(233)·$10^{-5}$ | $A_1$ | 4.011(110)·$10^{-5}$ |
| $\tau_1$ [ps] | 2.764(312) | $\tau_1$ [ps] | 2.391(172) |
| $A_2$ | 3.391(78)·$10^{-5}$ | $A_2$ | - |
| $\tau_2$ [ps] | 23.391(1.820) | $\tau_2$ [ps] | - |
| $t_0$ [ps] | -0.056(0.093) | $t_0$ [ps] | -0,015(0.040)·$10^{-2}$ |
| $y_0$ | 4.812(51)·$10^{-5}$ | $y_0$ | 3.391(35)·$10^{-5}$ |
| FWHM[a] [ps] | 0.273(50) | FWHM[a] [ps] | 0.305(84) |

[a] Post-fitting refinement

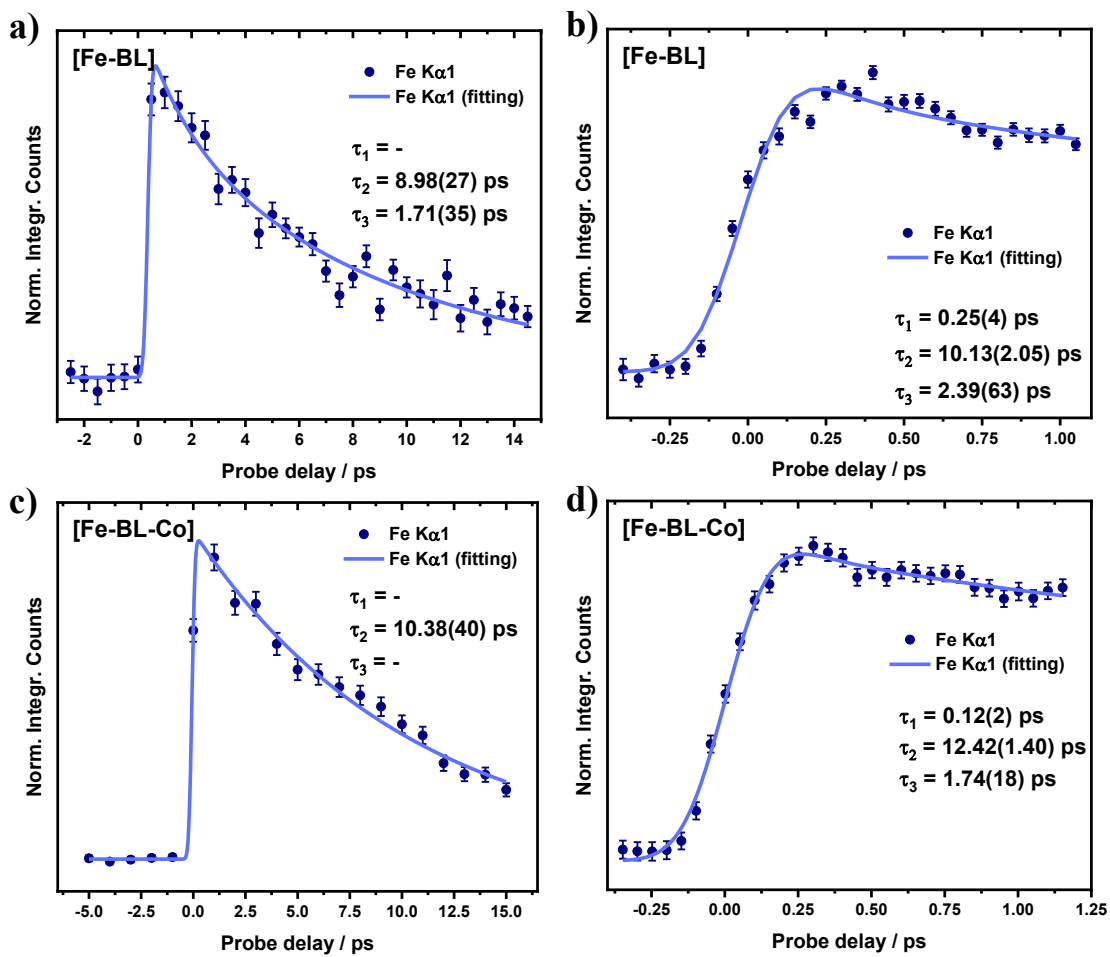

**Fig. S3.3.** Fluorescence decay fitting results for: **a)** [Fe-BL], long-time window; **b)** [Fe-BL], short-time window; **c)** Fe @ [Fe-BL-Co], long-time window; **d)** Fe @ [Fe-BL-Co], short-time window.

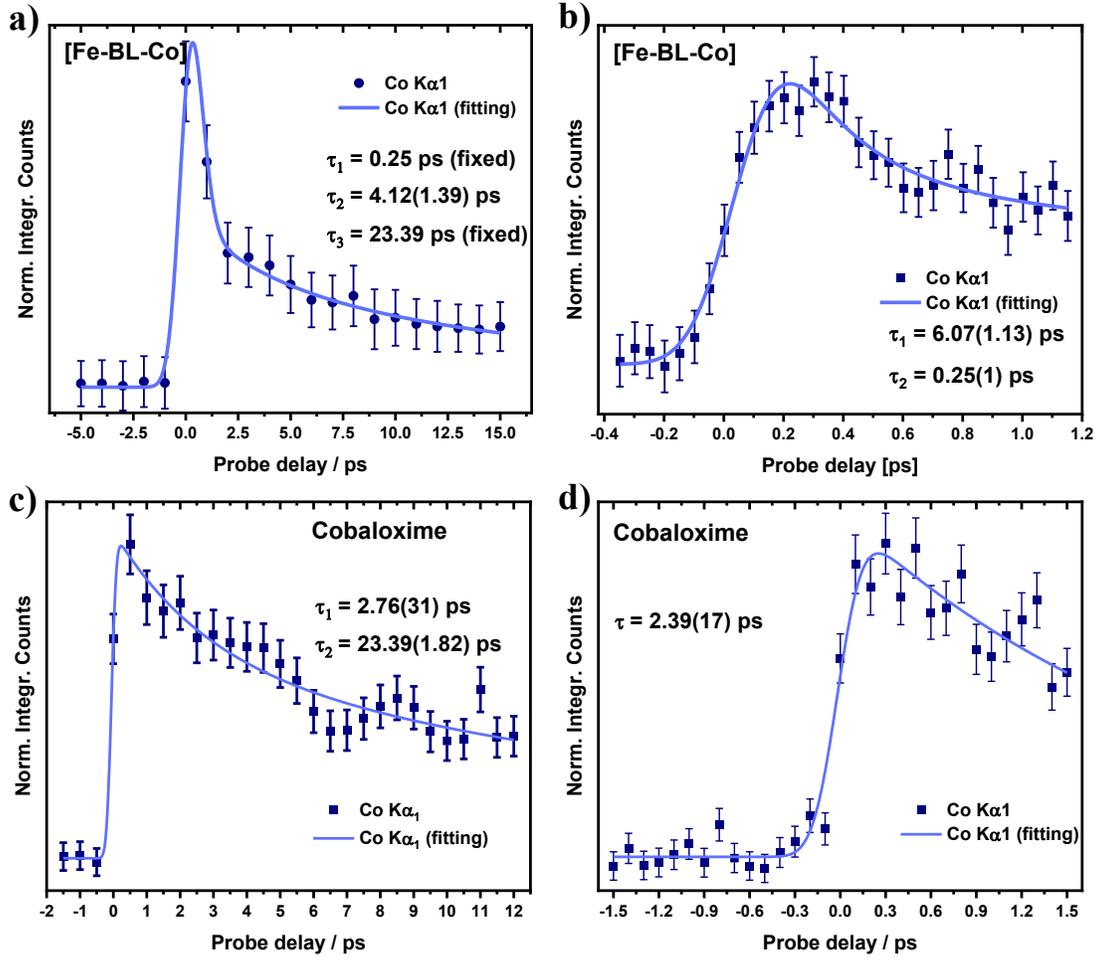

**Fig. S3.4.** Fluorescence decay fitting results for: **a)** [Co] @ [Fe-BL-Co], long-time window; **b)** [Co] @ [Fe-BL-Co], short-time window; **c)** cobaloxime, long-time window; **d)** cobaloxime, short-time window.

b) A direct and non-direct contribution to Kα XES

At the optical pump wavelength of 400 nm both Fe and Co centres were simultaneously excited (although predominantly Fe site) and probed with 9.3 keV X-rays. Hence, we have carried out rigorous and detailed analysis to distinguish the direct and non-direct contributions originating from the photoexcitation at the Co center in the studied dyad. In this case, we have computed the X-ray and optical convoluted cross-section relation $C_\sigma$ of cobaloxime and Co part of the dyad as follows:

$$C_\sigma \sim \left. \sigma_{X-Ray}^{cobaloxime} \sigma_{UV-VIS}^{cobaloxime} \middle/ (\sigma_{X-Ray}^{dyad} - \sigma_{X-Ray}^{PS})(\sigma_{UV-VIS}^{dyad} - \sigma_{UV-VIS}^{PS}) \right. \approx 0.59$$

This number can be compared to the average ratio between the intensity of cobaloxime kinetic trace and [Co] part of dyad kinetic trace, which was estimated to be 0.88. Our analysis yields that approximately 59% of Co signal in dyad originates from the different electronic structure around Co site in the dyad, as compared to isolated cobaloxime. The theoretical cross sections were computed using values listed in the NIST database. Optical cross sections were taken from UV-Vis spectra shown in Fig. 1 b.

### c) Nuclear wavepacket analysis

The oscillatory signals, previously extracted from the kinetic traces of both [Fe-BL] and [Fe-BL-Co] shown in Fig. 4 of the main text, were analyzed in detail using the Fourier transform analysis and are presented in Fig S3.4.

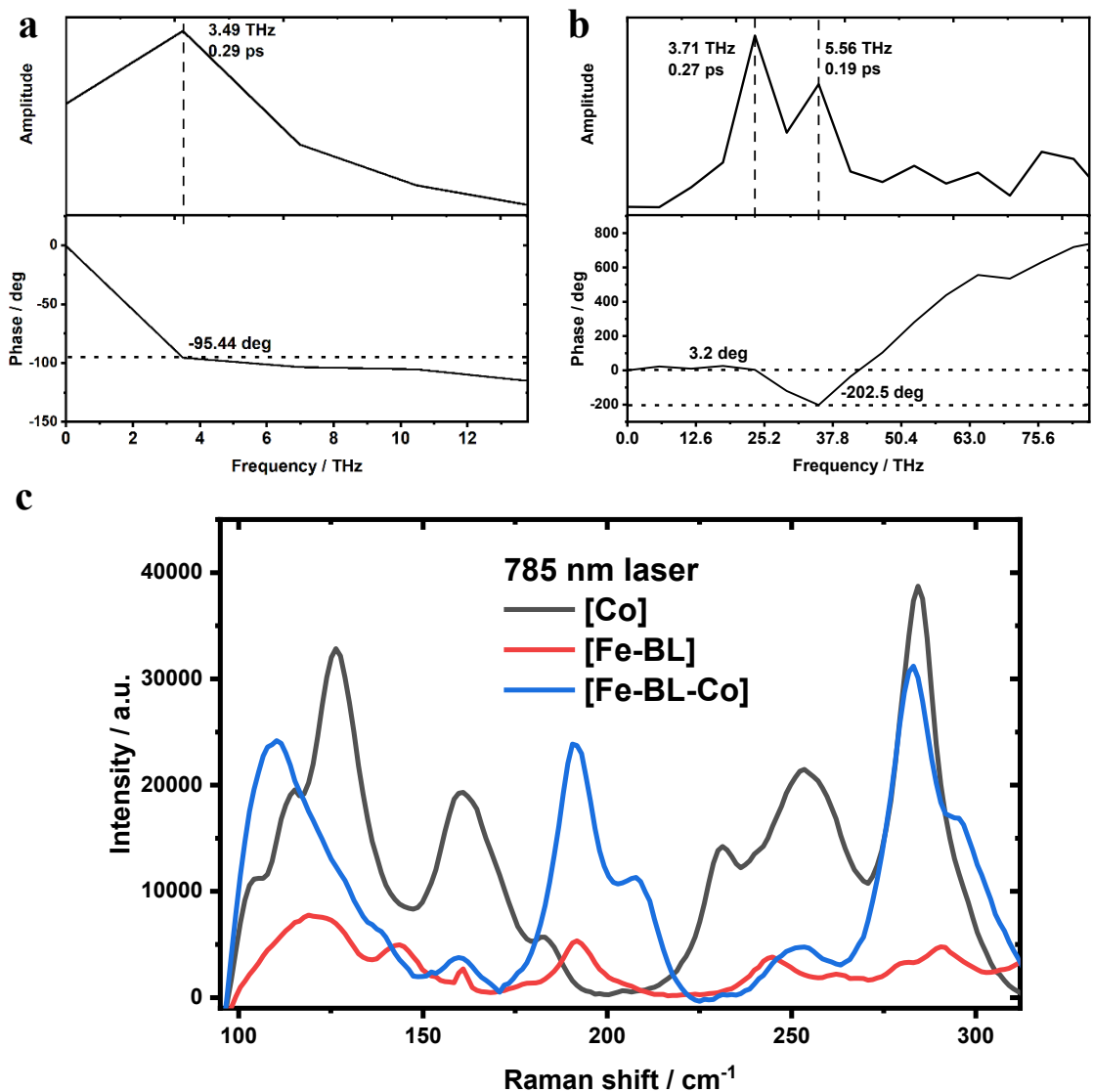

**Fig. S3.5.** Fourier transform of the oscillatory parts: **a)** [Fe-BL]; **b)** Fe part of [Fe-BL-Co]. Raman spectra of [Co], [Fe-BL], and [Fe-BL-Co] obtained with 785 nm laser (**c**).

For [Fe-BL], we found one dominating frequency of 3.49 THz (the corresponding half-period of 0.29 ps) and for the [Fe-BL-Co], there are two dominating frequencies of 3.71 THz (0.27 ps) and 5.56 THz (0.19 ps). These frequencies correspond to vibrational modes of 116.4 cm$^{-1}$ (for [Fe-BL]), 123.8 cm$^{-1}$ and 185.5 cm$^{-1}$ (for [Fe-BL-Co]), respectively, in the range of one degree of freedom for single bond thermal oscillations at 25 °C ($k_bT$). Noteworthy, similar frequencies could be resolved experimentally by means of Raman spectroscopy (see spectra in Fig. S3.4c). Specifically, the difference between 116.4 cm$^{-1}$ and 123.8 cm$^{-1}$ (~1 meV) could be observed only in terms of the amplitude, while a band around 185.5 cm$^{-1}$ should appear only in [Fe-BL-Co]. It must be underlined, that in TR-XES such differences would be difficult to be detected. Since the phase of Fourier transform can be affected by the presence of a significant noise contribution, we decided to use an additional method of analysis. The nuclear wavepacket motion parameters for [Fe-BL-Co] were refined with the use of the damped oscillatory, function:

$$f(t) = y_0 + e^{-t/t_0}\left(b_1 sin\left(\pi \frac{t-t_{c1}}{w_1}\right) + b_2 sin\left(\pi \frac{t-t_{c2}}{w_2}\right)\right) \qquad (S3.c.1)$$

where:

$y_0$ – intensity offset in a.u.; $t_{c1}$, $t_{c2}$ – phase shifts in ps; $w_1$, $w_2$ – oscillation period in ps; $t_0$ – damping factor in ps; $b_1$, $b_2$ – initial amplitude of oscillations in a.u.

The results are presented below. In case of [Fe-BL], the $b_2$ was set to 0. It is worth to mention that for [Fe-BL-Co] the damping factor is much bigger, although the uncertainty of this parameter prevents to derive any quantitative conclusions. Interestingly, the oscillation in [Fe-BL] changes phase upon cobalt coordination. This may suggest, that after forming the dyad, the longer oscillation is partially quenched and re-induced upon photoexcitation. The phase of oscillation in [Fe-BL] is ~ 9°, i.e. close to sinusoidal, therefore induced by impulse-stimulated Raman scattering.[16–18] On the other hand, the 0.26 ps component phase in [Fe-BL-Co] is ~51°, thus of mixed sine/cosine character and for 0.19 ps, the phase with ~89° is of cosine character. The cosine type of oscillation was proven to be a direct marker of the influence of the excited state generation upon photoexcitation and followed by coherent vibrations generation [18,19]. Thus at least one of the vibrations in [Fe-BL-Co] is related to the charge transfer accompanied by the Fe-ligand bond stretching.

**Table S3.2. Wavepacket analysis fitting results with equation (S3.c.1).**

| [Fe-BL] | | [Fe-BL-Co] | |
|---|---|---|---|
| $b_1$ | 0.064(176) | $b_1$ | -0.008(4) |
| $t_{c1}$ [ps] | 0.046(17) | $t_{c1}$ [ps] | 2.632(275) |
| $w_1$ [ps] | 0.284(22) | $w_1$ [ps] | 0.255(30) |
| $y_0$ | 0.003(4) | $b_2$ | 0.015(4) |
| $t_0$ [ps] | 0.402(150) | $t_{c2}$ [ps] | 8.425(370) |
| | | $w_2$ [ps] | 0.185(8) |
| | | $y_0$ | 0.000(2) |
| | | $t_0$ [ps] | 1.632(1.270) |

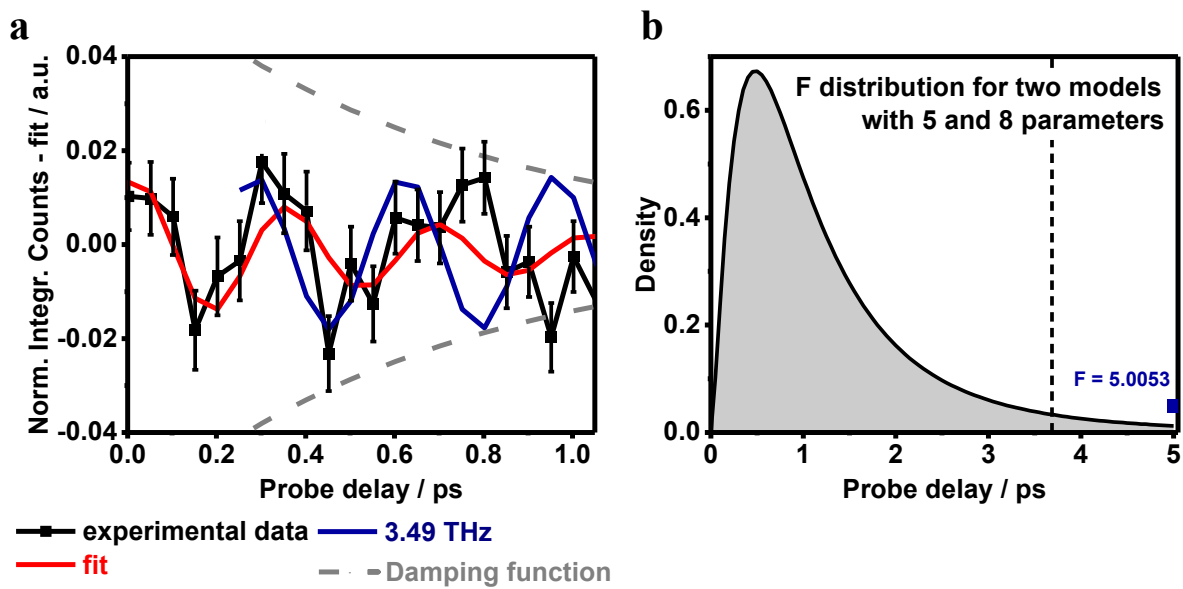

**Fig. S3.6. a)** Single oscillatory fit to wavepacket signal; **b)** F distribution (black) for nested 5-parameter model in 8-parameter model along with F-value in this study (blue). The vertical line indicates critical F-value of 3.6875 above which the null hypothesis can be rejected in the current conditions at p level of 0.05.

Given the large experimental error bars, we wanted to discard the possibility that the additional oscillation observed for the dyad could compensate for the high damping value. In other words, we have also verified the scenario, in which the extracted oscillations could be described with a single sine function. These fit results are shown in Fig. S3.5a and compared statistically with the double sine fit, previously presented in Fig. 4a, using an F-test (Fig. S3.5b). The F-test is intended to compare two models, where one of them contains less parameters and is nested into more complex one. The result of the test indicates statistical significance of the more complex fit, especially when $\chi^2$ comparison is not enough, thus preventing us from overfitting the data. In our specific case, the single-sine model was described by 5 parameters and was nested into a more complex model described with 8 parameters. The fitting procedure was carried out with experimental error bars as weights, and in F-value calculation, $\chi^2$ values of 0.01338 and 0.00814 were found for single- and double-sine fit functions, respectively. Since all fits were done with p = 0.05, the null hypothesis in this test stated that fitted functions are not different in 95% probability. For such conditions, the critical F-value to reject null at 95% probability was 3.6875, while the obtained value was 5.0053. Therefore, we can conclude that both models are statistically different and thus confirm the presence of the second oscillation in [Fe-BL-Co] kinetic traces.

### d) Co Kα₁ kinetic signals for -5-15 ps time window

The statistics of the Co Kα₁ long-time kinetic signal in [Fe-BL-Co] is substantially worse than for pure cobaloxime measurement, owing mainly to the two factors. First of all, due to substantially different absorption cross sections in the UV-Vis range, upon the photoexcitation of the dyad, the [Fe-BL] part is predominantly excited, while the direct excitation of the Co part is nearly completely avoided. Secondly, the X-ray beam intensity is distributed over two metal centers. Since with the measurement of cobaloxime alone, the aim was to detect excited states formed upon direct photoexcitation, and no [Fe-BL] was present. Still, the long kinetic trace for [Fe-BL-Co] substantially differs in shape as compared to pure cobaloxime, especially in the initial 5-6 ps range (sec. 3a, Fig. S3.3 a, and c). Consequently, although a possible long-lived component in Co moiety could not be excluded, it is visible only after the first ~7 ps of the [Fe-BL-Co] Co Kα₁ kinetics, where the signal almost reaches the background level. It would not affect the presence of $\tau_{1,FeCo}$, and short-time kinetic trace, since contributions from the shortest time constants appear at the beginning of the kinetic evolution. Moreover, the step size in this measurement was equal to 1 ps, which corresponds to approx. 2 data points that represent time constant of 1 ps. The initial fitting results of the kinetic traces for Co Kα fluorescence in [Fe-BL-Co] in 15 ps time window are shown in Fig. S3.3 a. All fit parameters were left as free and as a result we obtained good fit results with 2 time constants of 1.4 and 17 ps, respectively. The result is shown in Fig. S3.5 and the corresponding parameters are summarized in Table S3.3 (column A). Interestingly, the fitting of Co Kα fluorescence to the 1.2 ps kinetic trace revealed unambiguous presence of another ultrashort contribution of 0.25 ps. The 1 ps time constant in the 15 ps kinetic fit, was statistically represented by a single point, therefore it could be an artefact. To verify this, we repeated the fitting of the Co Kα fluorescence to the 15 ps kinetic trace using a fixed time constant of 0.25 ps and keeping all other fit conditions the same. However, the fit did not converge to any reasonable result, and therefore we extended the fit model with an additional time constant. First, we fitted all 3 decays, and the fitting procedure produced a very large time constant and high uncertainties. This contribution was interpreted as a representation of an electronic state with decay significantly longer than the time window of measurement. Therefore, the large value was fixed, and data were re-fitted. The result is shown in Fig. S3.5 b and corresponding parameters are available in Table S3.3 under column B. The results again exhibit 1 ps time constant (and did not require the shortest 0.25 ps time constant), which was interpreted as an artefact and a time constant of 7 ps with a very high uncertainty. For purpose of testing this hypothesis further, we employed a fit procedure with 3 time constants, of which two: 0.25 ps and infinite were fixed. The result is shown in Fig. S3.5 c and corresponding parameters are available in Table S3.3 under column C. The results concluded that the 1 ps value from the first fitting attempt was indeed an artefact due to too low temporal step size of the measurement. Earlier reports suggest that a two-exponential, sequential decay for cobaloxime (one fast and second around 20 ps), and we assumed the same scenario for our longer time constants.[20] Therefore, $A_2$, $\tau_2$, $A_3$, and $\tau_3$ represented a decay of the LMCT state to the ground state through the MC state upon direct photoexcitation, while $\tau_1$ and $A_1$ represented M'MCT transition. This implied, that the corresponding amplitudes of the $A_2$ and $A_3$ in the model from Table S3.3 C will be similar, because both concern excited states in the same simple decay pathway. However, the difference between them is around a factor of 2. Due to the fact, that the LMCT excitation is still present, we assumed that the related decay pathway will correspond to the one in an isolated cobaloxime complex, especially for the lowest-lying state. A final fitting attempt was conducted on the model described in Table S3.3 C, with $\tau_3$ fixed at the value obtained from cobaloxime, namely 23.39 ps. The result is presented in Fig. S3.3A and Table S3.1. Notably, the relation between $A_2$ and $A_3$ is almost 1. The time constant of 4.1 ps is discussed in the main text. The value of $\tau_3$ was later re-evaluated with other time constants fixed, and a value of 29.39(14.46) ps was obtained.

**Table S3.3. Co Kα fluorescence decay fitting results for [Fe-BL-Co] with different assumptions.**

| A | | B | | C | |
|---|---|---|---|---|---|
| $A_1$ | 4.453(1.535)·10⁻⁴ | $A_1$ | 4.101(844)·10⁻⁴ | $A_1$ | 3.160(116)·10⁻³ |
| $\tau_1$ [ps] | 1.36(25) | $\tau_1$ [ps] | 1.16(50) | $\tau_1$ [ps] | 0.25 (fixed) |
| $A_2$ | 1.914(103)·10⁻⁴ | $A_2$ | 1.874(757)·10⁻⁴ | $A_2$ | 1.967(114)·10⁻⁴ |
| $\tau_2$ [ps] | 16.74(3.54) | $\tau_2$ [ps] | 7.11(5.67) | $\tau_2$ [ps] | 5.97(1.19) |
| $t_0$ [ps] | -0.56(7) | $A_3$ | 5.993(4.267)·10⁻⁵ | $A_3$ | 6.393 (117)·10⁻⁵ |
| $y_0$ | 1.041(48)·10⁻⁵ | $\tau_3$ [ps] | infinite (fixed) | $\tau_3$ [ps] | infinite (fixed) |
| FWHM$^a$ [ps] | 0.284 (fixed) | $t_0$ [ps] | -0.57(5) | $t_0$ [ps] | -0.01(2) |
| | | $y_0$ | 1.025(25)·10⁻⁴ | $y_0$ | 1.024(25)·10⁻⁴ |
| | | FWHM$^a$ [ps] | 0.284 (fixed) | FWHM$^a$ [ps] | 0.284 (fixed) |

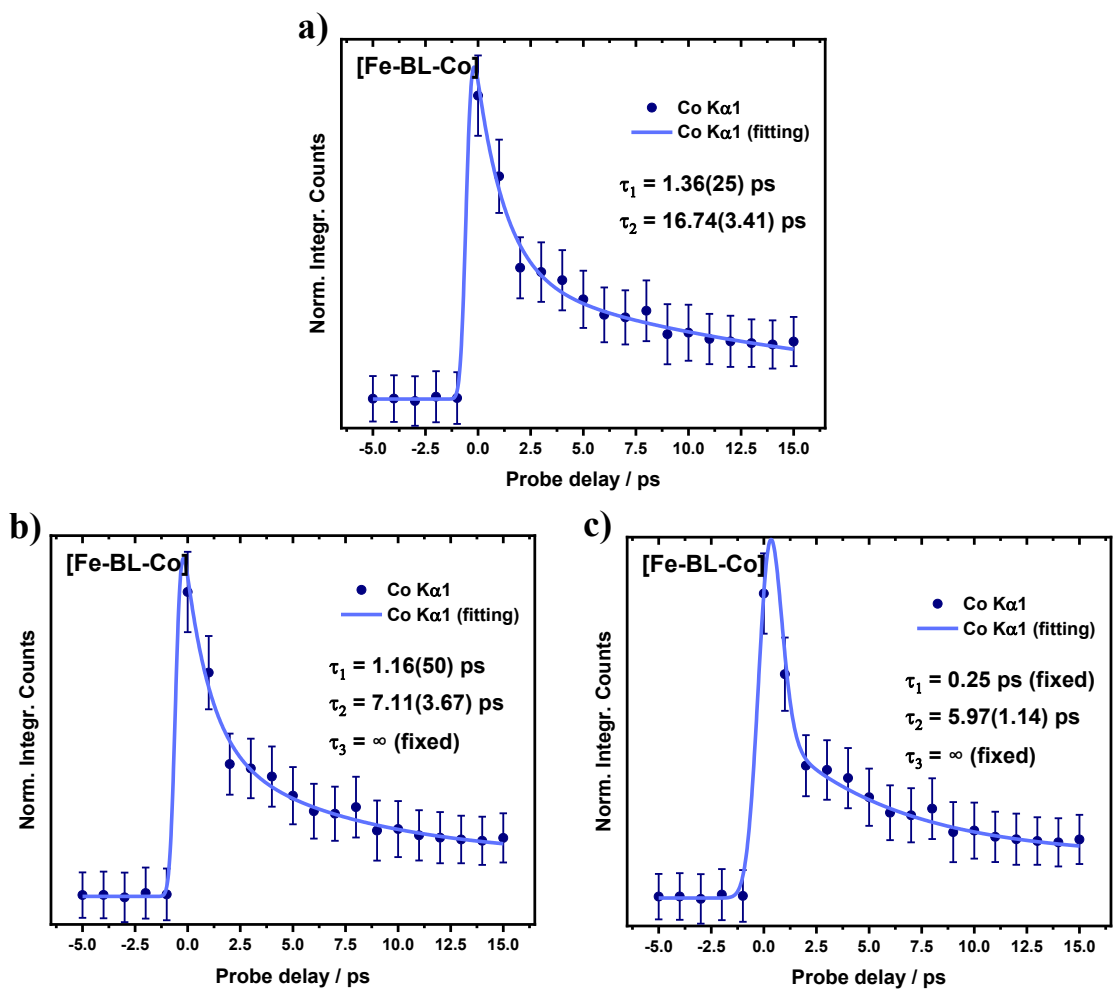

**Fig. S3.7.** Co Kα fluorescence decay fitting results for [Fe-BL-Co] 15 ps kinetics: a) with 2 time constants; b) with 3 time constants and very long time constant fixed; c) with 3 time constants and 0.25 ps and infinite time constants fixed.

### e) *d-d* interactions in Co

The *d-d* interactions can occur when different valence *d* orbitals in a metal complex are not fully occupied. In octahedral symmetry, the selection rules forbid *d-d* transitions to occur and thus they can be observed when the symmetry of the complex is significantly distorted, for example by the Jahn-Teller effect. Even though the symmetry of $Co(dmgH)_2Cl(py)$ complex is significantly distorted, thus allowing weak *d-d* transitions to occur, the UV-Vis spectrum of pure cobaloxime does not contain the characteristic *d-d* bands (Fig. 1 b) observed for another family of cobaloximes with axial alkyl and amino ligands $[Co(dmgH)_2(Alkyl)(Base)]$[21]. Therefore, the low-lying acceptor state for LMCT deexcitation is not populated due to the *d-d* transition. In order to assign the time constants obtained via the fitting of TR-XES kinetic traces to possible transitions in our Co complex, the $\tau_{1,Co}$ was tentatively assigned as LMCT → MC decay, while $\tau_{2,Co}$ represent the MC relaxation in this model. We want to underline, that the MC nature of the second state must be independently confirmed, yet the LMCT state was also confirmed by DFT results, making MC state an obvious candidate as an acceptor for the LMCT decay. The proposed state diagram for the relaxation on the Co site of the dyad is also analogical to numerous Fe polypyridyl complexes with long-lived MC states.[22,23] A dedicated Co Kβ TR-XES experiment would further confirm the nature of the assigned states involved in the decay process, which was beyond the scope of the present study.

### f) Kinetic model for XES fluorescence kinetic traces.

*Fe Kα XES.* For Fe in [Fe-BL-Co], two decay channels were proposed (Fig. 5):

1) alpha (α) channel: $^{1/3}\text{MLCT}^* \xrightarrow{k_1} {}^3\text{MLCT} \xrightarrow{k_2} {}^3\text{MC} \xrightarrow{k_3} \text{gs}$;

2) beta (β) channel: $^{1/3}\text{MLCT}^* \xrightarrow{k_4} {}^3\text{MC} \xrightarrow{k_3} \text{gs}$.

They were described by a system of differential equations:

$$\frac{d\,{}^1MLCT}{dt} = -k_1\,{}^1MLCT - k_4\,{}^1MLCT \tag{S3.1.a}$$

$$\frac{d\,{}^3MLCT}{dt} = k_1\,{}^1MLCT - k_2\,{}^3MLCT - k_{ct}\,{}^3MLCT \tag{S3.1.b}$$

$$\frac{d\,{}^3MC}{dt} = k_4\,{}^1MLCT + k_2\,{}^3MLCT - k_3\,{}^3MC \tag{S3.1.c}$$

$$gs = M - {}^1MLCT - {}^3MLCT - {}^3MC \tag{S3.1.d}$$

$$^1MLCT(t=0) = M \tag{S3.1.e}$$

$$^3MLCT(t=0) = 0 \tag{S3.1.f}$$

$$^3MC(t=0) = 0 \tag{S3.1.g}$$

where: M – initial excited state fraction. For the modelling purposes, an equal relation between concentration and signal strength was assumed. The system of differential equations above was solved numerically in Mathematica 11 software and all solutions were broadened by Heaviside step function under the convoluted with normalized Gaussian function to model rise time of electronic state:

$$g_{broadened}(t) = \frac{1}{2\sigma\sqrt{2\pi}} \int e^{-\frac{y^2}{2\sigma^2}} h(t - t_0 - y) g(t) dy \tag{S3.3}$$

where:

$g(t)$ – broadened function defined by one of eq. S3.1.a – S3.1.g;

$h(t - t_0 - y)$ – Heaviside step function;

$\sigma$ – Gaussian standard deviation function.

The final fitted function was as follows:

$$f_{Fe}(t) = y_0 + {}^1MLCT + {}^3MLCT + {}^3MC \tag{S3.4}$$

where:

$y_0$ – vertical offset.

*Co Kα XES.* There were two decay paths identified in [Co]:

1) M'MCT $\xrightarrow{k_7}$ gs;

2) LMCT $\xrightarrow{k_5}$ MC $\xrightarrow{k_6}$ gs.

The charge transfer (CT) was treated as instantaneous, therefore it is completed within the IRF function. The M'MCT state was acting as an acceptor of CT from bridging ligand BL, while LMCT state was representing direct optical excitation and was described by an independent $k_5$ rate constant. The analogical transition was observed in pure cobaloxime kinetic data with 2.76 ps time constant. The differential formula system with boundary conditions was as follows:

$$\frac{dM'MCT}{dt} = -k_7 M'MCT \tag{S3.5.a}$$

$$\frac{dLMCT}{dt} = -k_5 LMCT \tag{S3.5.b}$$

$$\frac{dMC}{dt} = k_5 LMCT - k_6 MC \tag{S3.5.c}$$

$$gs_{Co} = M'MCT_0 + LMCT_0 - M'MCT - LMCT \tag{S3.5.d}$$

$$M'MCT(t=0) = M'MCT_0 \tag{S3.5.e}$$

$$LMCT(t=0) = LMCT_0 \tag{S3.5.f}$$

The final fitting function was:

$$f_{Co}(t) = y_0 + M'MCT + LMCT + MC \tag{S3.6}$$

The $M'MCT$ and $LMCT$ were also broadened by the function described in eq. 3.

All fitting results are summarized in Table S3.4. In the first approach, decay constant values obtained from the fluorescence decay formula fitting were fixed. The amplitudes and offsets were fitted. Afterward obtained values were fixed to refine the rate constants.

**Table S3.4.** Summary for fitting of kinetic equations to experimental data.

| Fe in [Fe-BL] // Fe in [Fe-BL-Co] | | | [Co] in [Fe-BL-Co] dyad // cobaloxime | | |
|---|---|---|---|---|---|
| M | 0.736(19) | 0.608(9) | M'MCT $^0$ | 0.966(160) | - |
|  |  |  | LMCT $^0$ | 0.730(64) | 0.925(68) |
| $t_0$ [ps] | -0.034(8) | 0.002(0.0005) | $t_0$ [ps] | -0.005(0.015) | -0.018(0.021) |
| $y_0$ | 0.047(14) | 0.027(0.007) | $y_0$ | 0.002(0.032) | 0.004(0.048) |
| $k_1 / \tau_1$ [ps$^{-1}$/ps] | 4.527(1.495) / 0.221(72) | 4.527 (fixed) / 0.221 (fixed) | $k_5 / \tau_5$ [ps$^{-1}$/ps] | 4.12 (fixed) | 2.76 (fixed) |
| $k_{et} / \tau_{et}$ [ps$^{-1}$/ps] | - | -0.294(35) / 3.404(405) | $k_6 / \tau_6$ [ps$^{-1}$/ps] | 3.659(1.246) / 0.273(93) | - |

The $k_2$, $k_3$, $k_4$, and $k_5$ rate constants were calculated from $k = 1/\tau$ relation. Decay time constants $\tau$ were taken from fluorescence fitting results. In total, five parameters were fitted to the experimental data: M, $y_0$, $t_0$, $k_1$ and $k_{et}$. Unlike for the FWHM value used in the fluorescence fitting, the σ parameter here is a standard deviation of Gaussian in the IRF function. The re-evaluated value of $\sigma = 0.106(5)$ ps corresponds to 0.25(1) ps from the fluorescence fitting, which clearly resembles the FWHM of IRF function (0.28 ps). The $t_0$ was set as a free parameter to ensure fit convergence. The $k_1$ decay rate represents a $^1$MLCT->$^3$MLCT transition, and the inclusion of this decay rate is necessary due to nuclear wavepacket motion shown in Fig. 4 and Fig. S3.4. According to DFT results, eventual charge transfer should go through BL ligand acceptor state. The CT in the [Fe-BL] part of the dyad was considered to originate from two possible states: the $^3MLCT$ or $^3MLCT$ . In both cases, to model CT, an additional element of $-k_{ct}\,^3MLCT^*$ or $-k_{ct}MLCT$ was added into differential equations S3.1.b and S3.1.a respectively. In the first case, any fitting attempts were unsuccessful. For CT from $^3$MLCT state, a two-step analysis was applied. First, for [Fe-BL] a rate constant $k_1$ for $^1$MLCT→ $^3$MLCT transition was evaluated. For CT from $^3$MLCT in [Fe-BL-Co] this value was fixed, and the fit was conducted with the free $k_{ct}$ parameter. Results for global kinetic fitting are presented in Figs. 5, S3.6 and S3.7.

In the case of cobaloxime analysis, a two-state decay model was used to match the fluorescence decay results. Two scenarios were included, where states either decay in parallel or in a hierarchical way. Only the hierarchical model was reproducing the data with the states diagram (Fig. S3.8) and fitted kinetic traces presented in Fig. S3.7. In early kinetics (-1-1 ps), the signal dynamics can be described by 2.76 decay time, while a longer period requires the inclusion of a second, infinite time constant. This model was the basis for [Fe-BL-Co] analysis with 4.12 ps time fixed in a short-time window (Fig. S3.6). Importantly, the dynamics in short- and long-time windows could be reproduced with 2 kinetic constants. The results of kinetic equation fitting (Fig 5 c, Fig. S3.6 b) confirmed the fluorescence kinetic trace analysis for Co in [Fe-BL-Co]. For the sake of precision, two models were also tested: the M'MCT and LMCT states decaying in a hierarchical and parallel way. Only the model presented in Fig. 5 a reproduced data with satisfactory quality. Due to a very low Co signal intensity in [Fe-BL-Co] the estimated uncertainties are high.

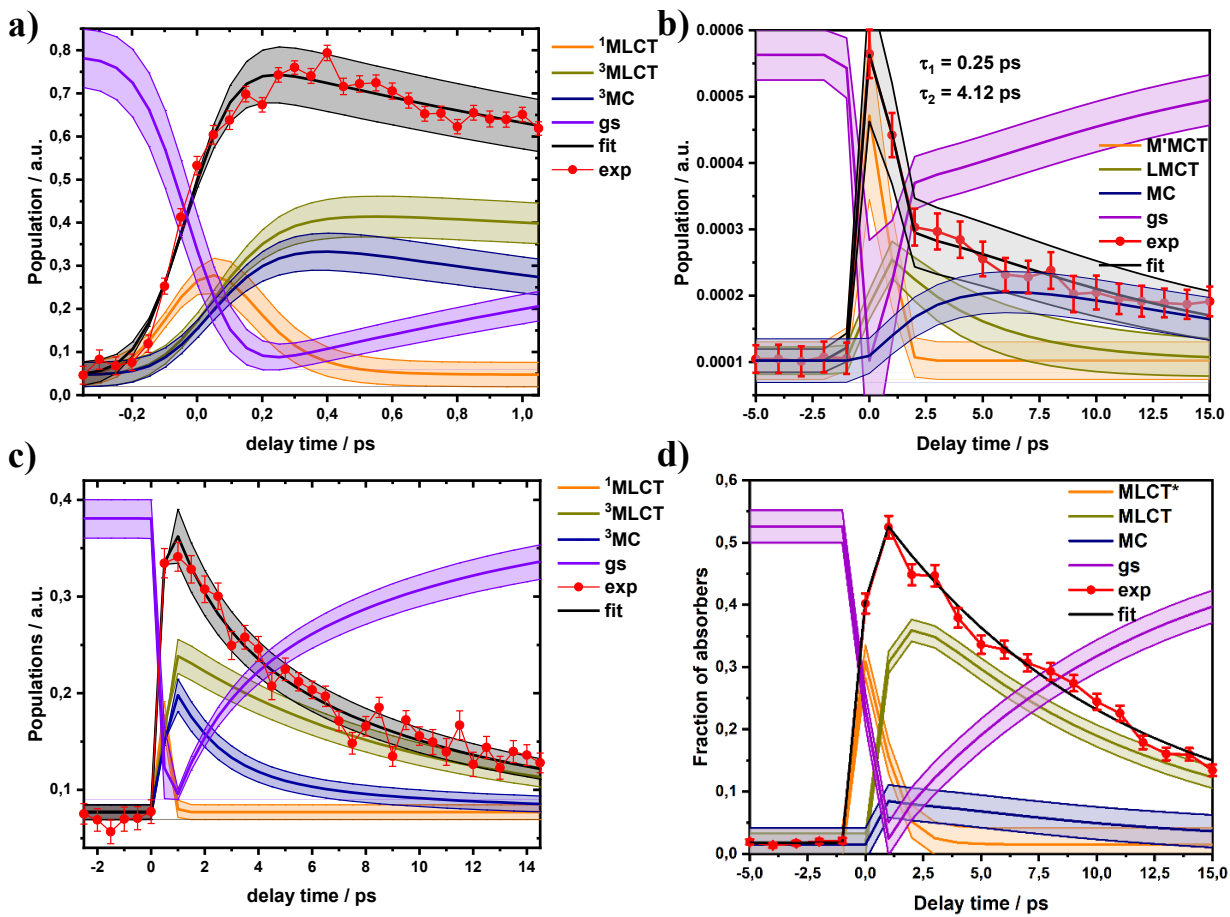

**Fig. S3.8.** Total kinetic traces for: **a)** [Fe-BL], short-time window; **b)** Co @ [Fe-BL-Co], long-time window; **c)** [Fe-BL], long-time window; **d)** Fe @ [Fe-BL-Co], long-time window.

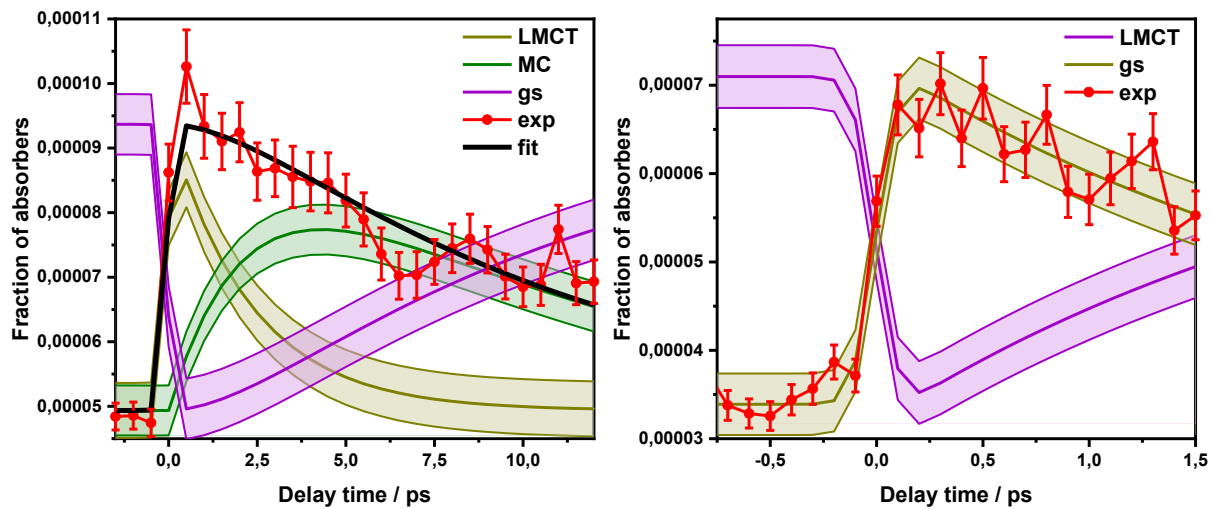

**Fig. S3.9.** Populations for short- and long-time window kinetic traces for cobaloxime: **a)** long-time window; **b)** short-time window.

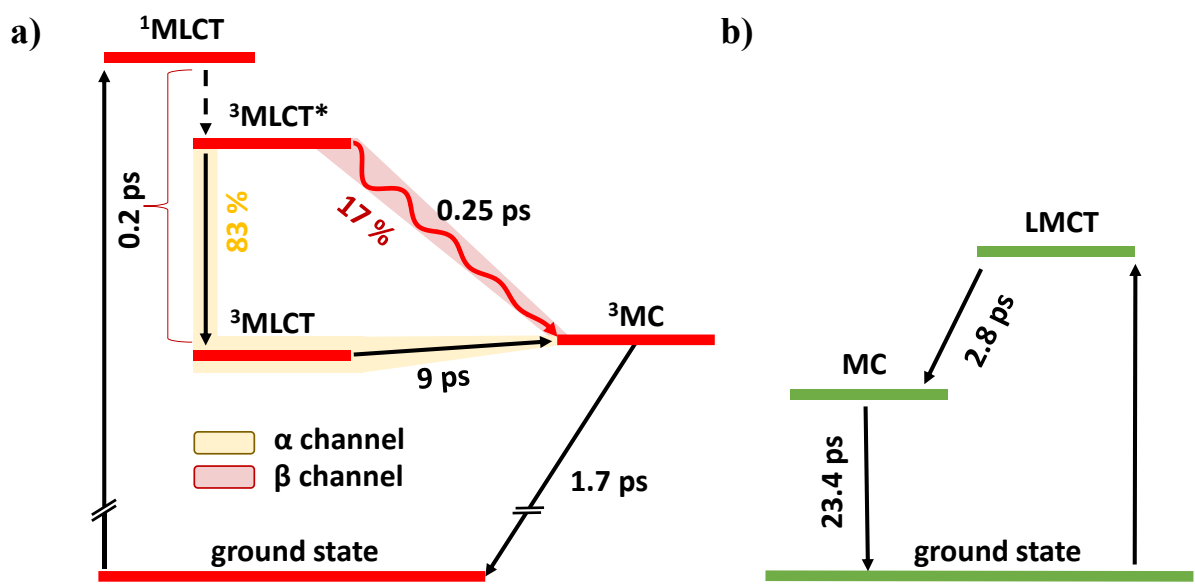

**Fig. S3.10.** State diagram for: **a)** [Fe-BL] and **b)** cobaloxime.